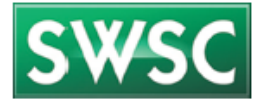

RESEARCH ARTICLE                                                To Be Published As Open ∂ Access

# Space Climate and Space Weather over the past 400 years: 1. The Power input to the Magnetosphere

Mike Lockwood [1,*], Mathew J. Owens [1], Luke A. Barnard [1], Chris J. Scott [1], and Clare. E. Watt [1]

[1] Department of Meteorology, University of Reading, Earley Gate, Reading, RG6 6BB, UK
[*] Corresponding author: m.lockwood@reading.ac.uk

ABSTRACT

Using information on geomagnetic activity, sunspot numbers and cosmogenic isotopes, supported by historic eclipse images and in conjunction with models, it has been possible to reconstruct annual means of solar wind speed and number density and heliospheric magnetic field (HMF) intensity since 1611, when telescopic observations of sunspots began. These models are developed and tuned using data recorded by near-Earth interplanetary spacecraft and by solar magnetograms over the past 53 years. In this paper, we use these reconstructions to quantify power input into the magnetosphere over the past 400 years. For each year, both the annual mean power input is computed and its distribution in daily means. This is possible because the distribution of daily values divided by the annual mean is shown to maintain the same lognormal form with a constant variance. This study is another important step towards the development of a physics-based, long-term climatology of space weather conditions.

Key words. Magnetosphere – Space environment – Interplanetary medium – Historical records – Energy deposition

## 1. Historical data

In recent years, reconstruction of space climate conditions (in the form of annual means of relevant parameters), has been made possible from five main types of historic observation series relating to some aspect of solar magnetism. These records measure different aspects of space climate and extend back into the past over intervals of different lengths.

### 1.1 Telescopic Sunspot Observations

When, where, and by whom the telescope was invented is a matter of debate because it was preceded by a number of simpler optical devices (King, 1955; Van Helden, 1977; Watson, 2004). However, there is little doubt that the realisation of the telescope's application spread from the Dutch town of Middelburg, the site of a glass factory using Italian glass-making techniques. The invention was claimed by Jacob Metius of Alkmaar and Sacharias Janssen of Middelburg, but all we know for sure is that a spectacles manufacturer in Middelburg, Hans Lippershey, filed for a patent in 1608. Following reports of this patent, scientists around Europe began experimenting with the device in 1609, including Thomas Harriot in England and Paolo Sarpi in Italy. It is probable that it was through his friend and patron, Sarpi, that Galileo Galilei came to hear of the device and his work greatly improved its design. Johannes Kepler in Prague was able to borrow one of Galileo's telescopes from Duke Ernest of Cologne and, in improving it further, he founded the science of optics in 1610.

Naked-eye observations of sunspots had been made since ancient times, many by Chinese and Korean astronomers, but only when atmospheric conditions were favourable (e.g., Willis et al., 1980) and when the political situation meant they were sought by rulers as a source of guidance (e.g., Hayakawa et al., 2015). The first known telescopic observations were by Thomas Herriot towards the end of 1610, as recorded in his notebooks (North, 1974), followed shortly after by those by Christoph Scheiner and Johannes Fabricus in March 1611 (Vaquero and Vázquez, 2009). Fabricus had been a medical student and returned home from University in Leiden with some telescopes which, with his father, he began to use to observe the Sun: he was the first to publish such observations in his 22-page pamphlet "De Maculis in Sole Observatis" ("On the Spots Observed in the Sun") which appeared in the autumn of 1611. Galileo started his scientific study of spots in April 1612 after he received a copy of Scheiner's written report and published his first reports in 1613. Telescopic observations of sunspots have been made ever since and composite records compiled to give a near-continuous record over 400 years. Calibration of different observers that contributed to this record is very difficult and composites vary in detail, whilst agreeing on the major features (Clette et al., 2014, Lockwood et al., 2016a, Usoskin, 2017). In particular, "daisy-chaining" of sequences using intercalibration observers from intervals of simultaneous observations is fraught with difficulties and potential errors (Lockwood et al., 2016b) and techniques that calibrate all observers to a common standard are greatly preferable (Usoskin et al., 2016) but can only be applied to data for after 1750. In recent years there has been an ongoing effort to improve the composites by improving the intercalibrations between different observers and to lower the weighting given to data that are not of sufficient quality (Clette et al., 2014, Usoskin, 2017).

1.2 Geomagnetic Observations

Whereas regular observations of sunspots were established in just a few short years after the invention of the telescope, the next major development took over a century to come to fruition. It was made by George Graham in London, Alexander von Humboldt in various locations, Olof Peter Hiorter in Uppsala and Carl Friedrich Gauss in Göttingen. Graham first reported geomagnetic activity in 1724, Hortier noted its regular diurnal variation in 1740 and the connection of large transient events to aurora in 1741 (a magnetic disturbance also seen by Graham in London). From his geomagnetic observations at various locations, von Humboldt introduced the concept of a "geomagnetic storm" and sparked the interest of his friend Gauss in 1828. Gauss subsequently developed the first magnetometer that could reliably measure the field strength and/or its horizontal component, establishing the first geomagnetic observatory in Göttingen in 1832. Reviews of the development of the observation of geomagnetic activity have been given by Stern (2002) and Lockwood (2013). Some composites have used geomagnetic activity data from soon after this date; for example, Svalgaard and Cliver (2010) used regressions with different types of geomagnetic data to extend the sequence back to 1831. However, there are concerns about the calibration, stability and homogeneity of these earliest data and so Lockwood et al. (2014) consider the usable data to start in 1845.

1.3 In-situ observations

A key development was the advent of observations of Earth's magnetosphere and near-Earth interplanetary space by spacecraft. The solar wind had been postulated to occur in transient bursts which caused geomagnetic storms by Chapman and Ferraro (1931) but was deduced to be continuous from studies of comet tails by Cuno Hoffmeister in 1943 and Ludwig Biermann in 1951 (see Biermann et al., 1967). Initial in-situ indicators of the solar wind were reported from the Lunik 2 and 3 spacecraft in 1959, but the first unambiguous data were recorded in 1961 by Explorer 10. When Mariner 2 flew towards Venus in 1962 it found the solar wind to be continuous and also showed fast and slow streams that repeated with the rotation of the Sun. Parker (1958) predicted that this flow should bring with it a magnetic field of solar origin and this "interplanetary magnetic field" (IMF) was also detected by Explorer 10 and Mariner 2. The Explorer 10 data were initially a puzzle, being complicated by an orbit that meant the spacecraft continuously moved between the magnetosphere and the shocked solar wind of the magnetosheath. The Mariner2 magnetometer suffered from spacecraft magnetic cleanliness issues: hence the first reliable and continuous data on the IMF were obtained only when IMP-1 was launched towards the end of 1963 (see review by Parker, 2001). A full survey of the heliospheric field at all heliographic latitudes (the IMF being the heliospheric field in the ecliptic plane) only commenced in 1992 when the Ulysses spacecraft used the gravitational field of Jupiter to move out of the ecliptic plane (Balogh et al., 1992; see review by Owens and Forsyth, 2013).

Observations of the solar wind and the embedded IMF have been made since 1963, and collected into the much-used "OMNI" dataset by the Space Physics Data Facility at NASA/Goddard Space Flight Center (Couzens and King, 1986). Observations are relatively sparse until 1966 and do not become (almost) continuous until the advent of the WIND and ACE spacecraft in 1996 (Finch and Lockwood, 2007).

1.4 Solar Magnetograph data

In 1904 George Ellery Hale discovered magnetic fields in sunspots using the Snow telescope and an 18-foot focal length solar spectroscope at Mount Wilson Observatory (MWO) from the Zeeman splitting of spectral emission lines, confirmed in 1908 using the 60-foot "solar tower" constructed the year before (Hale, 1908). However, full-disk magnetograph observations did not begin until 1957 after Horace Babcock had developed the principles of the modern magnetograph (Babcock, 1953) and a version of his instrument was installed at the 150-foot tower at MWO. Routine full disk measurements of the Sun's photospheric field (magnetograms) began at MWO in 1966, and similar measurements began both at the National Solar Observatory on Kitt Peak (NSO) and Wilcox Solar Observatory (WSO) roughly a solar cycle later in the mid-1970s (Stenflo, 2015).

The Potential Field Source Surface (PFSS) method to model the solar corona from magnetograms was developed by Schatten et al. (1969) and Altschuler and Newkirk (1969). In this method, photospheric magnetic fields are mapped to the top of the solar corona on the basis of number of assumptions. This involves solving Laplace's equation within an annular volume above the photosphere in terms of a spherical harmonic expansion, the coefficients of which are derived from Carrington maps of the photospheric magnetic field (i.e., magnetograms assembled over an entire solar rotation from magnetographs on Earth's surface or on board a spacecraft in orbit around the L1Lagrange point). Two major assumptions are that temporal variations within the 27 days taken to build up the map have no effect and that there are no current sheets in the corona (these are neglected so as to allow unique solutions in closed form). To eliminate the possibility that such simple harmonic expansions would result in all of the magnetic field lines returning to the Sun within a small heliospheric distance, the coronal field was required to become radial at the outer boundary, termed the "source surface". Essentially PFSS yields the equilibrium (minimum energy) configuration that the corona is heading towards for a given distribution of photospheric field, whereas the state of the real corona at any time is also dependent on its time history. Despite its many assumptions and obvious limitations, PFSS has been very successful in the study of a wide range of solar and heliospheric phenomena, (see review by Mackay and Yeates, 2012) and, in the present context, has supplied information about the tilt of the Heliospheric Current Sheet (HCS) (see review by Owens and Forsyth, 2013). PFSS has been applied routinely to WSO magnetograms since 1976

1.5 Cosmogenic isotopes

In addition to these four sets of observations, there is one other dataset which both extends these data sequences and helps bring them together into a coherent understanding. Cosmogenic isotopes are products of Galactic Cosmic Rays (GCRs). They tell us about the state of the Sun because the GCR flux is modulated by the heliosphere and, in particular, the HMF dragged out of the Sun because it is embedded in the solar wind flow. Three isotopes have been particularly valuable in this context: $^{10}$Be (mainly measured in ice sheets), $^{14}$C (mainly measured in tree rings) and $^{44}$Ti (found in meteorites) (see review by Usoskin, 2013). The $^{10}$Be and $^{44}$Ti are spallation products (respectively, from impacts on O and N atoms in the atmosphere and Fe and Ni atoms in the meteorite. The variations in the $^{14}$C isotope production are much larger because they arise from capture of thermal neutrons that are generated in large numbers by relatively low energy secondary particles. These provide proxy datasets which, unlike the other "as-it-happened" observations, can be improved, for example, by taking more and/or better measurements of the terrestrial reservoir where the isotopes are deposited, and by improved modelling of the production of the isotopes and of their transport and deposition through the Earth system. These records extend back over thousands of years but their indirect nature means that there is need to use both heliospheric and Earth-system models to interpret them. However, they can be used to check and confirm variations of reconstructed parameters such as the near-Earth HMF (the IMF, Owens et al., 2016a;b) and the calibrations used in composite data series, such as sunspot number (Asvestari et al., 2017).

1.6 Derived Open Solar Flux

Figure 1 summarises some annual means from these data and compares the durations of the datasets. So that they can be compared, each has been used to generate the first of the space-climate parameters to be reconstructed, namely the signed open solar flux (OSF, $F_S$), as discussed in the next section. This is the total flux of one polarity leaving the top of the solar corona. Figure 1a shows the open solar flux from two sources. The red line shows $F_S$ derived from PFSS modelling based on a combination of MWO and WSO magnetograph data (Wang et al., 2000). The blue dots joined by the thin blue line give the value derived from the OMNI data from interplanetary spacecraft: these values are derived from the radial component of the IMF in the ecliptic plane which gives information about the HMF at other latitudes (and hence the total open flux) because of the result from the Ulysses observations that the radial component of the heliospheric field is independent of latitude (Smith and Balogh, 1995; Lockwood et al., 2004; Owens et al., 2008). These data have used the solar wind velocity observations to remove the contribution of "excess" or "folded" flux using the kinematic correction developed by Lockwood et al. (2009), but similar answers can be obtained from the IMF radial field alone if averaging is carried out over 1 day, which effectively removes most of the folded flux without averaging out genuine sector structure that maps back to the solar corona. Even and odd numbered solar cycles (from minimum to minimum) are shaded white and grey, respectively.

Figure 1b shows the geomagnetic reconstructions of $F_S$ by Lockwood et al. (2014a). The colours show the results from 4 different pairings of geomagnetic indices and the grey area the estimated 1-sigma uncertainties derived using Monte-Carlo fitting techniques where random noise is added to the data and each fit repeated 10,000 times. Figure 1c shows model fits to these reconstructions from geomagnetic data using the continuity equation model formulation of Owens and Lockwood (2012): for this modelling, the rate of OSF emergence is quantified using group sunspot numbers. Results are shown for 4 different composites of sunspot group numbers which are presented in figure 1e : (blue lines) the group sunspot number $R_G$ of Hoyt and Schatten (1998) (with various corrections and additions, detailed by Lockwood et al. (2014c) and without the 12.08 multiplication factor); (red lines) the corrected sunspot number, $R_C$ (again divided by the 12.08 factor so that it corresponds to a group number) of Lockwood et al. (2014b;c); (green lines) the "backbone" group number $R_{BB}$ of Svalgaard and Schatten (2016); and (black lines) a new composite $R_{14C}$ by that compares the 11-year running means of the sunspot data with radiocarbon abundance data to calibrate observers. This calibration cannot be continued past about 1930 due to the increasing anthropogenic influence on the isotope abundances after this date, it also has to make allowance for the variations in the cosmic ray shielding effect of the geomagnetic field on the production of $^{14}$C and its exchange between the atmosphere, oceans and biomass. We here use the decadal means of sunspot numbers derived from $^{14}$C abundance by Usoskin et al. (2014). These are the only sunspot number composites that extend back to before the Maunder minimum (shaded in pink in figures 1d and 1e).

Figure 1c shows that we cannot distinguish these sunspot number series using the geomagnetic reconstructions. A similar conclusion was reached by Lockwood et al. (2014b) and Lockwood et al. (2016a). The best fit variations of $F_S$ using these four sunspot group number data sequences are almost indistinguishable and all agree very well with the geomagnetic reconstructions, as well as the data from PFSS and the in-situ interplanetary spacecraft. The yellow band shows the ±1σ uncertainties computed from the fit residuals with the interplanetary data shown in figure 1a (and repeated in all panels of figure 1 that present $F_S$ data series).

However, if we extend these sunspot and modelled $F_S$ data series to before the start of reliable geomagnetic data (1845), and back to the start of regular telescopic monitoring of the Sun (1612), larger differences become apparent, as shown in figures 1d and 1e. There has been a great deal of recent literature about the composite sunspot number series (see reviews by Clette et al., 2014; Lockwood et al., 2016a; Usoskin, 2017). The corrections used in $R_C$ are simple and certainly not the last word on the subject; however, the similarity to $R_{14C}$ shows that, for the time being at least, using $R_C$ remains a valid option. There is now quite widespread agreement that $R_G$ is too small in the interval between the end of the Maunder minimum and the middle of the Dalton minimum (c. 1710-1810) (see reviews by, e.g., Clette et al., 2014 and Lockwood et al., 2016a) and this tendency can be seen in figure 1. $R_{BB}$ is intended as the replacement for $R_G$; however, there is concern that the daisy-chaining of calibrations inherent in $R_{BB}$ will compound errors as one goes back in time. In addition $R_{BB}$ assumes not only will the data from two observers vary linearly but that they are proportional, an assumption that other studies have shown to be invalid (Usoskin et al., 2016). Using the Royal Greenwich Observatory photoheliographic data and progressively degraded version versions thereof (degraded my omitting groups of area below a threshold value), Lockwood et al. (2016b) showed that this will cause sunspot numbers to be consistently overestimated as one goes back in time when observer acuity was lower because of instrumental limitations. This effect is consistent with the behaviour of $R_{BB}$ and the OSF modelled from it in figure 1.

Figure 1d also shows two OSF sequences derived from cosmogenic isotope records. The black line is the OSF modelled using $R_{14C}$, i.e. the sunspot number calibrated using the 14C radiocarbon data from tree rings around the world using the same model (i.e., that by Owens and Lockwood, 2012) and fitting procedure (minimisation of r.m.s. differences using the Nelder and Mead (1965) search algorithm) used to derive the variations from $R_C$, $R_G$ and $R_{BB}$. The cyan line is taken from annual values of the near-Earth interplanetary magnetic field, $B$, derived by McCracken and J. Beer (2015) from $^{10}$Be isotope abundances in the Dye 3 and the North Greenland Ice-core Project (NGRIP) ice sheet cores. We here use their data series in which potential major transient Solar Energetic Particle events have been identified from large increases in $^{10}$Be in a single year and removed by replacing with a linear interpolation from the surrounding years. We convert this estimate of the IMF $B$ to OSF $F_S$ using the polynomial fit to the geomagnetic reconstructions and near-Earth in-situ data given by Lockwood et al. (2014a). It should be noted that all data shown in figure 1(d), including the in-situ spacecraft data, have been passed through a 1-3-1 filter to smooth the great inter-annual variability in the $^{10}$Be data caused by the low numbers of the isotope atoms in the ice cores, often near the 1-count limit. As reported by Owens et al. (2016a; b), the reconstructions of $B$ based on geomagnetic and solar observations are very similar over the 1845-present interval; agreement of both of these series with cosmogenic-based reconstructions of $B$, is clearly present but is less close. Measurements of $^{44}$Ti in meteorites that have fallen to Earth since 1766 provide a valuable comparison with average levels of the early sunspot records. This proxy is affected by neither natural nor anthropogenic terrestrial influences, since the isotope is produced in space as a spallation product of Fe and Ni in the meteorite by irradiation by cosmic-ray protons (>70 MeV) during its path through the solar system to Earth. Because of its long half-life (59 years), $^{44}$Ti is insensitive to solar cycles but tells us about average levels on centennial scales. Asvestari et al. (2017) have shown that $R_C$ is the composite most consistent with the $^{44}$Ti measurements and that $R_{BB}$ is consistently too large. Note that, contrary to some recent suggestions, all the sequences in figure 1(d) indicate that the Maunder minimum is a deeper and longer-lived depression in sunspot activity than the Dalton minimum (c, 1800-1825) (see discussion by Usoskin et al., 2015).

## 2. Reconstructions of Space Climate

The basic technique facilitating reconstructions of average conditions in near-Earth space has been to take historic observations and interpret them using knowledge obtained from modern data (for example from spacecraft or magnetographs). Initially this was simply using single or multiple regression fits of co-incident data, but they have grown increasingly complex and intricate. Table 1 summarises the development of the reconstructions and models of past space climate.

Feynman and Crooker (1978) reconstructed annual means of the solar wind speed, $V_{SW}$, from the geomagnetic index aa, which extends back to 1868, using the fact that aa, like all "range" geomagnetic indices, has an approximately $V_{SW}^2$ dependence (see Lockwood, 2003). However on annual timescales, aa also has a dependence on the IMF field strength, *B*, which contributes considerably to the long term drift in aa. Lockwood et al. (1999) removed the dependence of aa on $V_{SW}$ using its 27-day recurrence (which varies with mean $V_{SW}$ on annual timescales) and computed the open solar flux (OSF, the total flux leaving the top of the solar corona) using "the Ulysses result" that the radial component of the heliospheric field is largely independent of heliographic latitude (Smith and Balogh, 1995; Lockwood et al., 2004; Owens et al., 2008). The principles of modelling this variation using the OSF continuity model were laid down by Solanki et al. (2000). They used sunspot number to quantify the OSF production rate, and although that parameterisation has become more sophisticated, something like it remains in use in all models today (Lockwood, 2003; Vieira and Solanki, 2010, Owens and Lockwood, 2012; Rahmanifard et al., 2017). The free parameters in the model were then defined by fitting to the geomagnetic reconstruction of Lockwood et al. (1999). The parameterisation of OSF production rate was improved by Solanki et al. (2002) and Vieira and Solanki (2010) and allowing calculation of photospheric flux which has, in turn, allowed solar irradiance modelling and comparisons with magnetograph data. The model of Owens and Lockwood (2012) developed the parameterisation of the OSF loss rate from a simple linear loss law to one that varies over the solar cycle, as predicted theoretically by Owens et al (2011). Because they are based on sunspot numbers, these models can make reconstructions that extend back to the start of regular telescopic observations of the Sun (*circa* 1612).

One point to note is that the continuity models apply to OSF but have also frequently applied to derive reconstructions of the near-Earth IMF (e.g. Rahmanifard et al., 2017) which requires understanding of how OSF and HMF are related: often the assumption is made, either explicitly or implicitly, that the two are linearly related (e.g., Svalgaard and Cliver, 2010). This assumption was indeed made in the analytic equations used in the first reconstruction of OSF by Lockwood et al. (1999) – however, this could be done only because the difference between the real OSF-HMF relation and the assumed linear one was then accounted for in the regressions then used to derive OSF from the data. However, in general for a fixed OSF, the near-Earth HMF will decrease with increasing solar wind speed because of the unwinding of the Parker spiral. In addition, as the solar wind velocity increases the longitudinal structure in the solar wind also increases (Lockwood et al., 1999) which means that the kinematic folding of open field lines is greater which increases the HMF for a given OSF. Hence these two effects are of opposite sense and have the same magnitude for near-Earth HMF of near 6nT: below this threshold the Parker spiral effect dominates and above it the flux folding effect dominates (Lockwood et al., 2009). The resulting relationship of OSF to near-Earth HMF has been studied by Lockwood and Owens (2011) and Lockwood et al. (2014a).

|  | **From geomagnetic data** | **Modelled using sunspot data** |
|---|---|---|
| Dates | ~1845-present | ~1640-present |
| Open Solar Flux (OSF), $F_S$ | Lockwood et al. (1999) | Solanki et al. (2000)<br>Lockwood (2003)<br>Owens and Lockwood (2012) |
| Photospheric flux |  | Solanki et al. (2002)<br>Vieira and Solanki (2010) |
| IMF field strength, *B* and Solar wind speed $V_{SW}$ | Svalgaard & Cliver (2005)<br>Rouillard et al. (2007)<br>Lockwood and Owens (2011)<br>Lockwood et al. (2014a) |  |
| Streamer Belt Width<br>OSF |  | Lockwood and Owens (2014) |
| IMF field strength, *B* | Owens et al. (2016a) | Owens et al. (2017)<br>Rahmanifard et al. (2017) |
| IMF field strength, *B* and Solar wind speed $V_{SW}$<br>Solar wind number density, $N_{SW}$ |  | Owens et al. (2017) |

**Table 1.** The development of reconstructions of annual means of near-Earth space conditions from observations of geomagnetic activity and from models using sunspot number observations as input

Svalgaard and Cliver (2005) noted that different geomagnetic indices have different dependencies on the IMF, *B* and the solar wind speed, $V_{SW}$, and therefore could be used in combination to derive both *B* and $V_{SW}$. The long-term variations that they derived have been questioned (see Lockwood et al., 2006) because they employed non-robust regression procedures to also filled large data gaps the IMF and solar wind speed time-series with interpolated values (a much more reliable option is to mask out the geomagnetic data during data gaps when the interplanetary data are missing, as used by Finch and Lockwood, 2007). However the insight of Svalgaard and Cliver (2005) is very valuable and Rouillard et al. (2009) exploited it to reconstruct both *B* and $V_{SW}$, and Lockwood (2014a) used 4 different pairings of indices to derive both with an uncertainty analysis back to 1845. There is always the possibility of calibration drift in the historic geomagnetic data (Svalgaard, 2014; Holappa and Mursula, 2015), some of these have been identified, agreed upon and corrected for (e.g. Lockwood et al., 2014d ) whereas others are still debated, and others noted but their effects not properly accounted for. This is an ongoing process which may possibly never arrive at a definitive conclusion (or if it does, it may not be clear that it has) for years in which data are sparse and/or of poor quality. However, there is a growing convergence between the different geomagnetic reconstructions of heliospheric parameters (Lockwood and Owens, 2011) and also with those from cosmogenic isotopes (Asvestari and Usoskin, 2016; Asvestari et al., 2016; 2017; Owens et al., 2016b) which indicates further corrections will be second order in nature, rather than fundamental.

As discussed above, the date to which we can extend back using the geomagnetic data is debatable and depends on an assessment at what date the surviving data can be considered sufficiently homogeneous, continuous and well-calibrated. Lockwood et al. (2014a) argue that this is 1845 but, for sure, it could not be before 1832 when Gauss introduced the first-properly calibrated magnetometer. Models based on sunspot number and the OSF continuity equation have been used to extend the sequence back from 1845 to the start of regular telescopic observations in 1612. Lockwood and Owens (2014) extended the OSF modelling to compute the flux in the streamer belt and in coronal holes and so computed the streamer belt width variation which matches well that deduced from historic eclipse images (Owens et al., 2017). The streamer belt width and OSF were used by Owens et al. (2017), along with 30 years of output from a data-constrained magnetohydrodynamic model of the solar corona based on magnetograph data, to reconstruct solar wind speed, number density and the IMF field strength based primarily on sunspot observations. Using these empirical relations, they produced the first quantitative estimate of global solar wind variations over the last 400 years and these are employed in the present paper.

The CMIP-6 project (the 6[th] Coupled Model Intercomparison Project) has led to the development of what are, to date, the most comprehensive and detailed set of solar forcing reconstructions for studies of global and regional climate and of space weather (Matthes et al., 2016). These extend back to 1850. The current paper is aimed at making reconstructions that extend back to close to the invention of the telescope (c. 1612), and so include the Maunder minimum, and also at including as much of the physics of solar wind generation and solar wind-magnetosphere coupling as possible – thereby making the extrapolation of conditions seen during the space age to quieter solar conditions more robust and also providing a better understanding of the implications of reconstructed annual mean values for the occurrence probabilities of space weather events of a given magnitude.

## 3. The need to use annual means

All the reconstructions discussed in section 2 are of annual resolution. The fundamental reason for this is that, through the process of magnetic reconnection in the dayside magnetopause, the coupling of solar wind energy, mass and momentum into the magnetosphere depends on the southward component of the IMF (in a frame relative to Earth's magnetic axis, such as Geocentric Solar Magnetospheric, GSM). Some of this variation is predictable, arising from the diurnal and annual variations in the Earth's magnetic dipole axis in a Sun-centred frame or from annual variations in the heliographic latitude of Earth (Lockwood et al., 2016). While these factors are predictable, others are not because much variation in IMF orientation in near-Earth space arises from stream-stream interactions in the heliosphere (of which Coronal Mass Ejections and Corotating Interaction Regions are the large end of the spatial spectrum), from Alfvén waves and from the solar wind's turbulent evolution from the solar corona to Earth (Lockwood et al., 2016c). We have no historic measurements from which we can separate these variations in IMF orientation from other variations in the solar wind. Hence the only way they can be dealt with is to average them out (Lockwood, 2013).

To illustrate how this works, we here make use of the IMF orientation factor in solar wind-magnetosphere coupling, $\sin^4(\theta/2)$ where the IMF "clock angle", $\theta = \tan^{-1}(|B_Y| / B_Z)$ and $B_Y$ and $B_Z$ are the Y- and Z-components of the IMF in the GSM reference frame. When the IMF component in the YZ plane points northward in the GSM frame, $\theta = 0$ and $\sin^4(\theta/2) = 0$; when it points southward in the GSM frame $\theta = \pi$ and $\sin^4(\theta/2) = 1$. In general, the function $\sin^4(\theta/2)$ is somewhat similar in effect to a half-wave rectified function such as $B_S/B$ where $B_S = -B_Z[GSM]$ when $B_Z[GSM] < 0$ and $B_S = 0$ when $B_Z[GSM] \geq 0$. However it is preferable because, unlike $B_S/B$, it has a continuous gradient as a function of clock angle $\theta$ and also because it does not give a sharp cut off in transfer to the magnetosphere when the IMF turns northward ($\theta$ falls below $\pi/2$). Figure 2 shows distributions of observed $\sin^4(\theta/2)$ for 1996-2016 (inclusive): this interval is used because the IMF data are almost continuous and data gaps are few and short. As a consequence, when we average over sub-intervals of duration $\tau$, the number of 1-minute samples in each sub-interval is essentially the same. The distributions are given for various timescales, $\tau$, over which the IMF data are averaged. In each panel, the red dashed line is the overall mean of 1-minute $\sin^4(\theta/2)$ values for the whole 21-year dataset, $<\sin^4(\theta/2)>_{all} = 0.355$. Because the numbers of samples in each averaging interval $\tau$ is the same, the "mean of the means" (sometimes called the "grand mean") equals $<\sin^4(\theta/2)>_{all}$, irrespective of the value of $\tau$ employed. Hence the red dashed lines give the mean of each of the distributions shown. The distribution y-scales are normalised with $N$ being the number of samples in each $\sin^4(\theta/2)$ bin, the largest value of which for a given $\tau$ is $N_{max}$. Panel (a) is for $\tau = 5$ min. Rounding up of the data to two decimal places artificially increases the occurrence of samples with $\sin^4(\theta/2) = 0$ and $\sin^4(\theta/2) = 1$, but only slightly. The distribution shows the full range of $\sin^4(\theta/2)$ factors from 0 to 1 and the peak of the distribution is at $\sin^4(\theta/2) = 0$ with no peak at or near the mean value $<\sin^4(\theta/2)>_{all}$. Part (b) is for $\tau = 1$ hr, and the averaging smooths out some of the variations but the full range of $\sin^4(\theta/2)$ factors is still seen and the peak is again at $\sin^4(\theta/2) = 0$ with no peak at or near $<\sin^4(\theta/2)>_{all}$. For $\tau = 6$ hrs (part c), the distribution is broad and asymmetric with a mode value that is increased but is still less than $<\sin^4(\theta/2)>_{all}$. Only for $\tau = 1$ day is the mode value approximately the same as $<\sin^4(\theta/2)>_{all}$ and the distribution close to symmetric. The standard deviation of the distribution for $\tau$ of 1 day is $\sigma = 0.42 \times <\sin^4(\theta/2)>_{all}$. This means that the error in applying the overall mean $<\sin^4(\theta/2)>_{all}$ to 1-day averages would give an error (at the 1-$\sigma$ level) of 42.0%. For $\tau = 27$ days (part e) the distribution is centred on $<\sin^4(\theta/2)>_{all}$ but narrower in width, such that this 1-$\sigma$ error is 10.3%, and for $\tau = 1$ yr (part f) this error is further reduced to 4.9%.

The reduction in the error caused by using the overall average value $<\sin^4(\theta/2)>_{all}$ with increasing $\tau$ is the main reason why annual means have been used in the reconstructions. Averaging over a year allows us to use $<\sin^4(\theta/2)>_{all}$, for each annual value to within a 1-sigma uncertainty of just 5% and hence the unknown effect introduced by the IMF orientation is averaged to the constant value of $<\sin^4(\theta/2)>_{all}$ to within this small error.

## 4. Power input into the magnetosphere

An equation to compute the power input from the solar wind into the magnetosphere, $P_\alpha$, was derived by Vasyluinas et al. (1982):

$$P_\alpha = (\pi L_o^2) \times (m_{sw} N_{sw} V_{sw}^3/2) \times (t_r) \qquad (1)$$

where $L_o$ is the mean radius of the magnetosphere as seen from the Sun, such that the magnetic field presents an area $\pi L_o^2$ to the solar wind flow and $(m_{sw} N_{sw} V_{sw}^2/2)$ is the dominant energy density in the solar wind, the bulk-flow kinetic energy density, so that $(\pi L_o^2) \times (m_{sw} N_{sw} V_{sw}^3/2)$ is the total energy flux incident on the magnetosphere ($m_{sw}$ is the mean ion mass, $N_{sw}$ the number density and $V_{sw}$ the speed). The term $t_r$ is a dimensionless transfer function, being the fraction of the incident energy flux that is transferred into the magnetosphere.

Assuming a hemispheric shape to the dayside magnetosphere, pressure balance at the nose of the magnetosphere gives

$$L_o = k_1 (M_E^2/P_{sw} \mu_o)^{1/6} \qquad (2)$$

where $k_1$ is a geometric factor for a blunt-nosed object, $M_E$ is Earth's magnetic dipole moment, $\mu_o$ is the magnetic constant, and $P_{sw}$ is the solar wind dynamic pressure, given by

$$P_{sw} = m_{sw} N_{sw} V_{sw}^2 \qquad (3)$$

Vasyluinas et al. (1982) adopted the dimensionless transfer function:

$$t_r = k_2 M_A^{-2\alpha} \sin^4(\theta/2) \qquad (4)$$

where $\alpha$ is the "coupling exponent" (a free fit parameter), $\theta$ is the clock angle that the IMF makes with the north in the Earth's GSM frame of reference (as discussed above) and $M_A$ is the Alfvén Mach number of the solar wind flow, related to $m_{sw}$, $N_{sw}$, $V_{sw}$ and the IMF field strength $B$ by

$$M_A = V_{SW} (\mu_o \, m_{sw} \, N_{sw})^{1/2} / B \qquad (5)$$

Substituting equations (2) - (5) into (1) yields

$$P\alpha = (k_1 k_2. \pi/2\mu_o^{(1/3-\alpha)}) \, M_E^{2/3} \, m_{sw}^{(2/3-\alpha)} \times B^{2\alpha} \, N_{sw}^{(2/3-\alpha)} \, V_{sw}^{(7/3-\alpha)} \sin^4(\theta/2) \qquad (6)$$

The interplanetary parameters $B$, $N_{sw}$ and $V_{sw}$ have been routinely measured by near-Earth, interplanetary craft. We here adopt a mean solar wind composition of 96% proton and 4% helium ions, giving $m_{sw} = 1.12$ a.m.u. The constant $(k_1 k_2)$ is here removed by normalising to the mean $P\alpha$ for a reference period, $Po$. This also removes any dependence on the assumed $m_{sw}$ - but this assumes that both $(k_1 k_2)$ and $m_{sw}$ remain constant. Note that the assumption of constant $m_{sw}$ is accurate to within a few percent. Consider the two dominant constituents of the thermal solar wind, namely alpha particles and protons:

using the near-continuous number density data for these ions from near the L1 Lagrange point over 1996-2016, $m_{sw}$ averages 1.102 a.m.u. and the standard deviation of 1-hour values is 0.062 a.m.u. and in annual values is 0.023 a.m.u., which gives errors in using the constant value at the 1-σ level of 5.63% and 2.09%, respectively. There is a clear but weak solar cycle variation in this $m_{sw}$ estimate, with the largest annual mean being 1.139 a.m.u. (a deviation from the mean of +3.4%) in the year 2000 (sunspot maximum) and the lowest value being 1.050 a.m.u. (a deviation of –4.7%) in the low sunspot minimum year of 2009.

The Earth's dipole moment varies on the centennial timescales of interest to this paper. We use the dipole moment variation provided by the gufm1 model prior to 1900 (Jackson et al., 2000) and by the IGRF-11 model after that date (Finlay et al., 2010): some intercalibration and smoothing is needed to prevent a discontinuity at 1900. One major and very important advantage of equation (6) is that there is just the one free fit parameter, α. It is possible, in principle, to include all of the factors in equation (6) with their own weighting factors and exponents and get extremely good fits that are, nevertheless, completely statistically meaningless and have no predictive capability. This pitfall is called "*over-fitting*" and it is a serious but under-appreciated problem in multiple regression analysis of geophysical time series that have internal "geophysical" noise. Overfitting refers to the situation when a fit has too many degrees of freedom and starts to approximate the noise in the training subset, which is not robust throughout the whole dataset. This is a recognised pitfall in areas where quasi-chaotic behaviours give large internal noise such as climate science and population growth (for example, Knutti et al., 2006; Knape and de Valpine, 2011) but is not recognised as widely in space physics where systems tend to be somewhat more deterministic with lower internal variability. The above theory constrains the relationship between the parameters, giving just one free fit parameter which greatly reduces the danger of over-fitting.

The coupling exponent α influences almost all the factors in equation (6) but is unknown and has to be derived empirically. The remainder of this section looks at best practice in deriving the optimum α for the annual mean data available from the reconstructions discussed in the previous section. It is shown that, although compiling $P\alpha$ from annual means of $N_{SW}$, $V_{SW}$, $B$ and $\sin^4(\theta/2)$ is a less satisfactory approach and does slightly alter the optimum α, it does not lower the correlation obtained, It is also shown that derivation of α is best done using the interplanetary data after 1996 which is continuous because data gaps have considerable influence.

To determine α, we need to assume a functional form for the dependence of a terrestrial space weather parameter or index on $P\alpha$. In general that form can be non-linear and is likely to be different for different space weather indices. We here demonstrate the principles using planetary geomagnetic index, Ap, which like all such "range" indices shows a linear dependence on $P\alpha$ on all timescales (Finch and Lockwood, 2007). This linearity will be demonstrated below.

The solid lines in Figure 3 show the linear correlation coefficients *r* between Ap and $P\alpha$, as a function of the α value used to compute Pα. These correlations are made using annual mean data but there is a complication here in that annual $P\alpha$ values can be constructed in two different ways. The simplest is to take annual means of variables, $M_E$, $B$, $N_{sw}$, $V_{sw}$ and $\sin^4(\theta/2)$ and then put these into equation (6). We here term results from such an "average-then-combine" procedure as $P\alpha_{ann}$. The second way is to take high resolution data (we here use 5-minute means), evaluate $P\alpha$ for each and then average them. We here term the results of such a "combine-then-average" procedure as $<P\alpha>_{\tau=1yr}$. The correlograms with Ap as a function of α are shown in figure 3 for $P\alpha_{ann}$ by the red solid line and for $<P\alpha>_{\tau=1yr}$ by the black solid line. The peak correlations for the two occur at quite similar α values (0.53 for $P\alpha_{ann}$ and 0.46 for $<P\alpha>_{\tau=1yr}$ marked, respectively, by the red and black vertical solid lines) . The peak is more marked for $<P\alpha>_{\tau=1yr}$ as *r* falls rapidly with α above the peak, whereas it falls only slowly for $P\alpha_{ann}$). Both constructions generate extremely high peak correlations with Ap ($r > 0.99$), being almost unity in both cases (details are given by cases A and B in Table 2). The fraction of the variation of Ap "explained" by a linear dependence on these $P\alpha$ estimates is $r^2$ and hence more than 98.6% of the variation in annual Ap values can be explained by the annual mean power input to the magnetosphere, as estimated by either procedure. In general, potential non-linearities make the combine-then-average procedure preferable, but because only annual mean data are available in reconstructions of past behaviour, the average-then-combine procedure is needed. The fact that that both give extremely high peak correlations at similar α values shows us that using annual means of interplanetary parameters is acceptable. Table 2 also gives the confidence level *S* of each correlation evaluated by comparison against the auto-regressive AR1 red noise model (thereby allowing for the "persistence" – also termed "conservation" or "auto-correlation" - in the two data series). In all cases *S* is less than 100% by only an extremely small amount.

|   | years (inclusive) | averaging interval, $\tau$ | number of samples, N | overall daily data availability | Ap series averaging | $P\alpha$ series | best-fit $\alpha$ | correlation coefficient, $r$ | confidence, $S$ (%) (to 3 decimal places) |
|---|---|---|---|---|---|---|---|---|---|
| A | 1996-2016 | 1 year | 21 | 98.60% | coincident data, $Ap_c$ | $<P\alpha>_{\tau=1yr}$ | 0.46 | 0.9931 | 100.000 |
| B | 1996-2016 | 1 year | 21 | 98.60% | coincident data, $Ap_c$ | $P\alpha_{ann}$ | 0.53 | 0.9942 | 100.000 |
| C | 1996-2016 | 1 day | 12198 | 98.60% | coincident data, $Ap_c$ | $<P\alpha>_{\tau=1day}$ | 0.46 | 0.9045 | 100.000 |
| D | 1964-2016 | 1 year | 53 | 63.01% | all data, Ap | $P\alpha_{ann}$ | 0.53 | 0.9213 | 99.998 |
| E | 1964-2016 | 1 year | 48 | 63.01% | coincident data, $Ap_c$ & years with $f < 0.15$ discarded | $<P\alpha>_{\tau=1yr}$, years with $f < 0.15$ discarded | 0.46 | 0.9932 | 100.000 |
| F | 1964-2016 | 1 year | 48 | 63.01% | coincident data, $Ap_c$ & years with $f < 0.15$ discarded | $P\alpha_{ann}$, years with $f < 0.15$ discarded | 0.53 | 0.9962 | 100.000 |

**Table 2.** Details of correlations between the Ap geomagnetic index and power input to the magnetosphere $P\alpha$.

The dashed lines in figure 3 show the corresponding correlograms for $P\alpha_{ann}$ and $<P\alpha>_{\tau=1yr}$ against the fraction of time in each year that Ap is in the top 5% of its overall distribution of values ($f$[Ap>Apo] where Apo = 36nT, the 95 percentile of the overall distribution of all daily means of Ap observations which began in 1932). The peak correlations are not as great as for $<Ap>_{\tau=1yr}$ but are still very high, exceeding 0.95, and are for almost the same $\alpha$ that gives peak correlations with the mean values. This fact will be exploited further in paper 2 (Lockwood et al., 2017b).

Figure 4 shows a scatter plot of the estimated normalised power input, $P\alpha/Po$, for the best-fit $\alpha$ against Ap for 1996-2016. The blue dots are for daily means $<P\alpha>_{\tau=1day}$ and $<Ap>_{\tau=1day}$ and use the best-fit $\alpha$ of 0.46. The best linear regression fit is the black line, given by $Ap_p = 13.32(P\alpha/Po)-0.43$. The correlation coefficient is 0.905 (case C in Table 2). The distribution of the fit residuals $[Ap-Ap_p]_{\tau=1day}$ is Gaussian (required for a valid linear regression) with a standard deviation of $\sigma = 5.4$nT. The red dots show the annual means with $P\alpha_{ann}/Po$ values computed from annual means of interplanetary parameters and using the best fit $\alpha$ of 0.53. The correlation coefficient is 0.994 (case B of Table A). The red line is the best fit linear regression to these points, $Ap_p = 12.32(P\alpha/Po)-0.01$. The distribution of the fit residuals $[Ap-Ap_p]_{\tau=1yr}$ is also Gaussian with a standard deviation of $\sigma = 1.1$nT.

The additional scatter in daily values is to be expected, physically as well as statistically, because the magnetosphere shows some persistence on timescales of several hours and so a daily value of Ap is expected to show some effect of energy input on the previous day (or maybe even a few days) as well as that on the day in question. This does not apply on annual timescales.

Figure 5 shows the annual-resolution time series giving the correlations listed in Table 2. Panels (a) and (b) are for the period of almost continuous data coverage (1996-2016), whereas (c) and (d) are for the full interval of interplanetary observations,1964-2016. Panels (b) and (d) give the fraction, $f$, of hours giving valid means of $P\alpha$ (requiring over 45 out of the 60 1-minute samples for each of $B$, $V_{SW}$, $N_{SW}$ and $\theta$ available in each year): (b) shows that $f$ is close to unity after 1996, but (d) shows that availability is very low for many of the years before that date. A threshold value of $f = 0.15$ is shown by the dashed line. In panel (a), the green line is the observed annual means of Ap, $<Ap>_{\tau=1yr}$, the black line is the linearly regressed variation of $<P\alpha>_{\tau=1yr}$ ($<P\alpha>'_{\tau=1yr}$, the prime denoting it has been linearly regressed with $<Ap>_{\tau=1yr}$) and the red line is $P\alpha'_{ann}$. The extremely close agreement between the three is quantified by the near-unity correlation coefficients for cases A and B in Table 2. Underneath the green line, the variation of Ap for only data that is coincident with a valid $P\alpha$ value, $<Ap_c>_{\tau=1yr}$, was also plotted; however, $f$ is so close to unity throughout this interval that it cannot be seen. The way $Ap_c$ was constructed was to linearly interpolate the 3-hourly observed values of Ap into hourly values and then all of those hourly values that were not coincident with a valid $P\alpha$ value (to within a propagation uncertainty of 15 min.) were removed ("masked out"). The remaining values were then averaged over the year to give $<Ap_c>_{\tau=1yr}$.

Although $<Ap>_{\tau=1yr}$ and $<Ap_c>_{\tau=1yr}$ are essentially identical in figure 5a, 5c shows this is far from true for several years before 1996. For example, between 1988 and 1996 there is a considerable difference and although the green line shows the true variation of Ap, the blue line (which we must correlate the means of $P\alpha$ with) is considerably different. Note that the means of <Ap$_c$>$_{\tau=1yr}$ and <P$\alpha$>'$_{\tau=1yr}$ (black line) are not shown for some years because *f* is too low to form a valid mean: the threshold availability for such means was set at *f* = 0.15, shown by the horizontal dashed line in 5d. The correlation between the blue and black lines in 5d, given by case E in Table 2, is exceptionally high. Repeating this correlation for the average-then-combine estimates, P$\alpha_{ann}$, very slightly increases the correlation, as it did for the post-1996 period (case F). Lastly the red line shows the results of correlating the P$\alpha_{ann}$ estimates with the Ap means, without first removing the non-coincident data (case D in Table 2). There are now enough years with *f* ≈ 1 to make the correlation very high but it is reduced and this is the only correlation that does not give a confidence level *S* that is not 100% to within 3 decimal places.

This analysis of the correlation of the power input to the magnetosphere, P$\alpha$, tuned with the coupling exponent $\alpha$ to give the best fit to the Ap geomagnetic data, shows that compiling P$\alpha$ from annual means of $N_{SW}$, $V_{SW}$, $B$ and $\theta$ does not lower the correlation obtained, although it is the less satisfactory approach. It does very slightly increase the optimum $\alpha$ from 0.46 to 0.53. This is important as it means we can use reconstructed annual means of these parameters to derive a valid reconstruction of annual means of P$\alpha$ (using $\alpha$ = 0.53). The analysis also shows that the correlations found for the near-continuous P$\alpha$ data after 1996 also apply to all the available data, provided data gaps in the P$\alpha$ data series are also masked out in the Ap data series so one compares like-with-like.

## 5. The relation of distributions of daily values to annual means.

Section 3 stresses how important averaging timescale $\tau$ can be by looking at the behaviour of the sin$^4(\theta/2)$ term in P$\alpha$. In this section we investigate the annual distributions of daily values ($\tau$ = 1day) of both Ap and P$\alpha$ and how they relate to the annual means ($\tau$ = 1 year).

The colour pixels in the middle panel of figure 6 show the annual probability density function (pdf, evaluated in Ap bins 1nT wide) of daily Ap values, <Ap>$_{\tau=1day}$, as vertical slices and as a function of year along the horizontal axis. The black line gives the annual means values, <Ap>$_{\tau=1yr}$ (shown by the green lines in figure 5). The fraction of time for which the Ap data are available (top panel) is unity for all years. The bottom panel shows the pdfs of <Ap>$_{\tau=1day}$/<Ap>$_{\tau=1yr}$ (evaluated in bins 0.1wide) . The black line shows the means of these normalised Ap distributions which, by definition, are always unity. What is striking is how uniform the distributions of <Ap>$_{\tau=1day}$/<Ap>$_{\tau=1yr}$ have been, despite the variation in <Ap>$_{\tau=1yr}$. There is a slight tendency for the mode value to increase slightly toward the mean value in the sunspot minimum years 1996 and 2009, but this is slight and overall these normalised distributions are remarkably similar. The implications of this will be discussed further and exploited in a second paper ("Paper 2").

This is of interest here as the good correlations between Ap and P$\alpha$ discussed in the previous section suggest the same should be true for the magnetospheric power input estimate P$\alpha$. This is investigated in figure 7, which is the same as figure 6 for P$\alpha$. The complication is that, unlike Ap, the data availability *f* is often much smaller than unity, and insufficient samples mean that the pdfs become very noisy. We here require that *f* (shown in the top panel), exceed 0.5 to construct a valid pdf. This is a compromise between the noise levels in the pdfs and the number of years available: the choice of *f* > 0.5 allows us to use data from 31 of the 53 years in the interval 1964-2016. The pdfs of < P$\alpha$ >$_{\tau=1day}$/Po are shown in the middle panel of figure 7 (evaluated for bins 0.1 wide) and those for < P$\alpha$ >$_{\tau=1day}$/< P$\alpha$ >$_{\tau=1yr}$ (also evaluated for bins 0.1 wide) in the bottom panel. Where sufficient data are available ( *f* > 0.5) the distributions of the normalised daily P$\alpha$ in the bottom panel of figure 7 are remarkably constant in form. Indeed they are more so than for Ap as even the slight increase in the mode value for 2009 cannot be detected, although it is (just) detectable for 1996.

The grey histograms in figure 8 present the overall distributions (all data from all years 1964-2016) of (a) <Ap>$_{\tau=1day}$/<Ap>$_{\tau=1yr}$ and (b) < P$\alpha$ >$_{\tau=1day}$/< P$\alpha$ >$_{\tau=1yr}$. In both cases a lognormal distribution has been fitted and is shown by the magenta line. By definition, both these distributions have arithmetic mean values of unity and so there is only one fit parameter, the variance, *v*. (Note that the mean and variance determine the other pair of logarithmic parameters, $\mu$ and $\sigma$, that can be used to define the lognormal distribution). It can be seen that both distributions are very close to lognormal. The distributions for individual years (the vertical slices in the bottom panels of figures 6 and 7) are very similar and give essentially the same variance *v* values as for the overall distributions. For Ap the overall distribution gives *v* = 0.934, whereas the mean *v* for the 53 years individually is 0.934±0.031 (the uncertainty being the standard deviation). For P$\alpha$, the overall distribution gives *v* = 0.913, whereas the mean *v* values for the 31 years with *f* > 0.5 is 0.915±0.025 (the uncertainty again being 1 standard deviation).

From this analysis we conclude that the distributions shown in figure 8 apply, to a good approximation, to all years individually. This has been shown to be true for the high < P$\alpha$ >$_{\tau=1yr}$/Po solar-maximum years, down to the lowest value seen since interplanetary measurement began, which was detected in 2009.

The near-constancy of the shape of the distributions of $<P\alpha>_{\tau=1day}/<P\alpha>_{\tau=1yr}$ has an interesting implication: variations in the annual means will be reflected in the variations in the numbers of days in the year that $P\alpha$ will exceed any fixed and large threshold that is used to define "large" events. The variation will not, in general, be linear but it will be monotonic. To test this implication, figure 9 looks at the variation of the fraction of days $f_p[ <P\alpha>_{\tau=1day} \geq P\alpha_{pc}]$ on which $<P\alpha>_{\tau=1day}$ exceeds or equals a threshold value $P\alpha_{pc}$. This is shown as a function of $<P\alpha>_{\tau=1yr}/Po$ for two fixed thresholds : $P\alpha_{pc} = 2.38Po$ (which defines the 95$^{th}$ percentile of the cumulative probability distribution of all available $<P\alpha>_{\tau=1day}$ values from the interval 1964-2016; and $P\alpha_{pc} = 3.77Po$ (which defines the 99$^{th}$ percentile of the same cdf). Only the 31years with data availability $f > 0.5$ from the interval 1964-2016 are used, which gives a total of 10186 daily means (out of the total available for the full interval of 12198). Figure 9 shows the annual fraction of daily mean $P\alpha$ values that are in the top 5% (solid points) and top 1% (open triangles) of all the 12198 available daily means. These data use the optimum $\alpha$ of 0.53. The solid line is a third order polynomial fit to the data points for $P\alpha_{pc} = 2.38Po$ and the dashed line is that for $P\alpha_{pc} = 3.77Po$. Both reveal that $f_p$ shows an increasing trend with the annual mean value, but it is not, in general, linear. The smaller numbers associated with higher percentiles of the cdf mean that the scatter is increased. The variations shown in figure 9 have an important implication because they mean that information about the annual averages, which can be reconstructed as described in section 2, gives us information about the number of days of high power input into the magnetosphere that leads to them being disturbed space weather days.

## 6. The variation of power input into the magnetosphere since 1612

The model reconstructions of Owens et al. (2017) give the annual mean values of $N_{SW}$, $V_{SW}$ and $B$ needed to evaluate $P\alpha$ from equation (6). Figure 2 shows that on annual time scales we can use a constant value of $<\sin^4(\theta/2)>_{all} = 0.355$ to an accuracy of a bit better than 5%. Other values are removed by normalising and we here divide by $Po$, the mean of $P\alpha$ for the period 1964-2016 (as used in previous sections). The Earth's magnetic moment $M_E$ is obtained from a spline of the IGRF and gufm1 models, as described earlier.

Figure 10 shows the time series of modelled annual values on near-Earth interplanetary parameters over 1612-2015 generated by Owens et al. (2017). Panel (a) shows the input sunspot number data series, $R_C$. The results do depend on which of the sunspot composite data series discussed in the introduction was adopted, but the study by Owens et al. (2017) shows that this sensitivity is not great.

The time series of annual $P\alpha$ computed from equation (6) with these model values depends on the coupling function $\alpha$ used. Much of the previous sections was based on $\alpha = 0.53$ as this gives $P\alpha$ that is proportional to Ap and predicts the Ap index to within 1 nT on annual timescales. However, this does not mean the same value of $\alpha$ would apply to all terrestrial disturbance indices as they will have different responses to a given energy input into the magnetosphere. Hence we here study the implications of a range of $\alpha$ values in figure 11.

Negative values of $\alpha$ have an obvious implication in that they generate solar cycle variations of $P\alpha$ that are in antiphase with the known solar cycles. Thus we can immediately discount their use as predictors of space weather conditions. Figure 11a shows the variation for $\alpha = 0$. From equation (6) this yields a $P\alpha$ that varies as $V_{sw}^{7/3}$ and $N_{sw}^{2/3}$, but is independent of the IMF $B$. This combination yields solar cycle variations that are almost flat, the main feature being a spike in the declining phase of the cycles when average solar wind speeds are enhanced by more frequent fast solar wind streams emanating from coronal holes at lower latitudes. The long-term variation is almost flat with only a slight depression during the Maunder minimum. Figure 11b is for $\alpha = 1/3$. From equation (6) this yields a $P\alpha$ that varies as $V_{sw}^2$, $N_{sw}^{1/3}$ and $B^{2/3}$. This yields small but recognisable solar cycles and a clear longer-term variation with a drift since the Maunder minimum that exceeds the amplitudes of recent solar cycles. This drift is clear in both sunspot minimum and sunspot maximum values. Figure 11c is for $\alpha = 2/3$ which yields a $P\alpha$ that varies as $V_{sw}^{5/3}$ and $B^{4/3}$ but is independent of $N_{sw}$. This yields a long-term drift in solar-cycle averages that is similar in magnitude to the amplitude of the recent cycles, but the drift in sunspot minimum values is smaller than that in sunspot maximum values. This dependence on $\alpha$ continues with increasing $\alpha$ as shown by figure 11d for $\alpha = 1$ (that gives variations with $V_{sw}^{4/3}$, $N_{sw}^{-1/3}$ and $B^2$), figure 11e for $\alpha = 4/3$ (that gives variations with $V_{sw}$, $N_{sw}^{-2/3}$ and $B^{8/3}$) and figure 11f for $\alpha = 5/3$ (that gives variations with $V_{sw}^{2/3}$, $N_{sw}^{-1}$ and $B^{10/3}$). In figure 11f, the variation is much greater in the solar maximum values than the solar minimum values, such that the overall variation rather resembles that in sunspot numbers. Note that for the larger $\alpha$ values, the power input decreases with increased $N_{sw}$ because the rise in energy density in the solar wind with $N_{sw}$ is smaller than the effect of reduced cross-sectional area of the magnetosphere intersecting the solar wind flow. These two effects balance at $\alpha = 2/3$ which is somewhat greater than all of the best-fit $\alpha$ values shown in Table 2. Hence although values are close enough to 2/3 to make the dependence on $N_{SW}$ weak, the dominant effect is that of the magnetospheric compression and energy input to the magnetosphere (weakly) decreases with increased solar wind number density.

The value of the coupling exponent derived here to fit Ap data of α = 0.53 gives factors of $B^{1.06}$, $N_{sw}^{0.14}$, and $V_{sw}^{1.80}$. This is very close to the dependence of $BV_{sw}^{n}$ found by Lockwood (2013) (without using $P\alpha$) for "range" geomagnetic indices, where *n* = 1.9 for the aa index, *n* = 1.8 for the Am index, and *n* = 1.6 for the Ap index. The black line in Figure 12 shows the centennial variation of $<P\alpha>_{\tau=1yr}/Po$ for α =0.53. This is superposed on coloured pixels giving the corresponding pdf of $<P\alpha>_{\tau=1day}/Po$ evaluated annually and in bins of 0.01. These are computed using the distribution of $<P\alpha>_{\tau=1day}/<P\alpha>_{\tau=1yr}$ shown in figure 8(b) which in section 5 was shown to be valid for individual years. The pdfs are evaluated using the fitted log-normal distribution for values of $<P\alpha>_{\tau=1day}/<P\alpha>_{\tau=1yr}$ that are $10^{-3}$ apart. These are then rescaled as $<P\alpha>_{\tau=1day}/Po$ values using the $<P\alpha>_{\tau=1yr}$ value for the year in question and then averages of the pdf taken over bins of $<P\alpha>_{\tau=1day}/Po$ that are 0.01 wide. The high preponderance of quiet days during deep solar activity minima when $<P\alpha>_{\tau=1yr}$ is low is shown by red and brown patches. The pale yellow pixels at high values show the variation of the probability of large space weather events when energy input into the magnetosphere is high.

## 6. Discussion and Conclusions

We have presented the first analysis of the variation of power input into Earth's magnetosphere over the last 400 years. This analysis has reconstructed not only the annual mean values, but also the annual probability distribution of daily means. Hence the reconstructions give both the average space climate, but also the probabilities of various levels of space weather conditions in the past.

The power input is not given here in absolute units of Watts, but as ratio of the overall mean for all observations in the space age (1964-2016), *Po*, partly because this removes the requirement to quantify the constants $k_1$ (the flow factor for a blunt-nosed object) and $k_2$ (the amplitude of the transfer function term), and partly because it cancels the dependence on the assumed constant average solar wind ion mass. We have shown how the derived $<P\alpha>/Po$ values depend on α but have concentrated on α = 0.53 which gives the best fit to the Ap geomagnetic index when annual mean data is used. Paper 2 in this series will carry this work forward and make reconstructions of geomagnetic activity indices and so look at the occurrence of geomagnetic storms and substorms over the past 400 years. We here note that the value of α = 0.53 is a close match to the dependence derived by Lockwood (2013) of a great many geomagnetic indices (specifically Ap, Kp, IHV, Am, AL and AE). These all show a dependence that is close to $BV_{SW}^2$. Other indices, such as Dst, IDV and m, depend almost solely on *B*.. Lockwood (2013) has explained the origin of the $V_{SW}^2$ term for the first of these two sets of indices as being due to their sensitivity to the substorm current wedge and predicts that the total current disrupted from the near-Earth cross-tail current and deflected down into the auroral electrojet in the current wedge will depend on the solar wind dynamic pressure and hence, at least approximately, on $V_{SW}^2$. Hence the expectation is that geomagnetic indices that respond strongly to substorm cycles of solar wind extraction of energy from the solar wind, its storage in the geomagnetic tail. and its subsequent energy deposition in upper atmospheric currents, will correlate well with $P\alpha$ for α somewhere between about 1/3 (which would give a $B^{2/3}V_{SW}^2$ dependence) and 1/2 (which would give a $BV_{SW}^{1.8}$ dependence). Paper 2 will study the long term variations in these geomagnetic indices, based on their high correlation with, and linear and proportional dependence on $P\alpha$. Other indices, in particular, the Dst index, do not show this proportionality with $P\alpha$ and the implication for Dst and ring current energetic particles will be discussed in later papers.

**Acknowledgments and Data** The authors are grateful to the staff of Space Physics Data Facility, NASA/Goddard Space Flight Centre, who prepared and made available the OMNI2 dataset used. The data were downloaded from http://omniweb.gsfc.nasa.gov/ow.html ). The geomagnetic data were obtained from a variety of data centres including the UK Solar System Data Centre at RAL (https://www.ukssdc.ac.uk/), The British Geological Survey (http://www.geomag.bgs.ac.uk/data_service/data/home.html), The World Data Centre WDC for Geomagnetism, Kyoto (http://wdc.kugi.kyoto-u.ac.jp/aedir/). The work of ML and MJO is supported by the SWIGS NERC Directed Highlight Topic Grant number NE/P016928/1 and the work of LAB, CEW and CJD by a STFC consolidated grant number ST/M000885/1. The editor thanks two anonymous referees for their assistance in evaluating this paper.

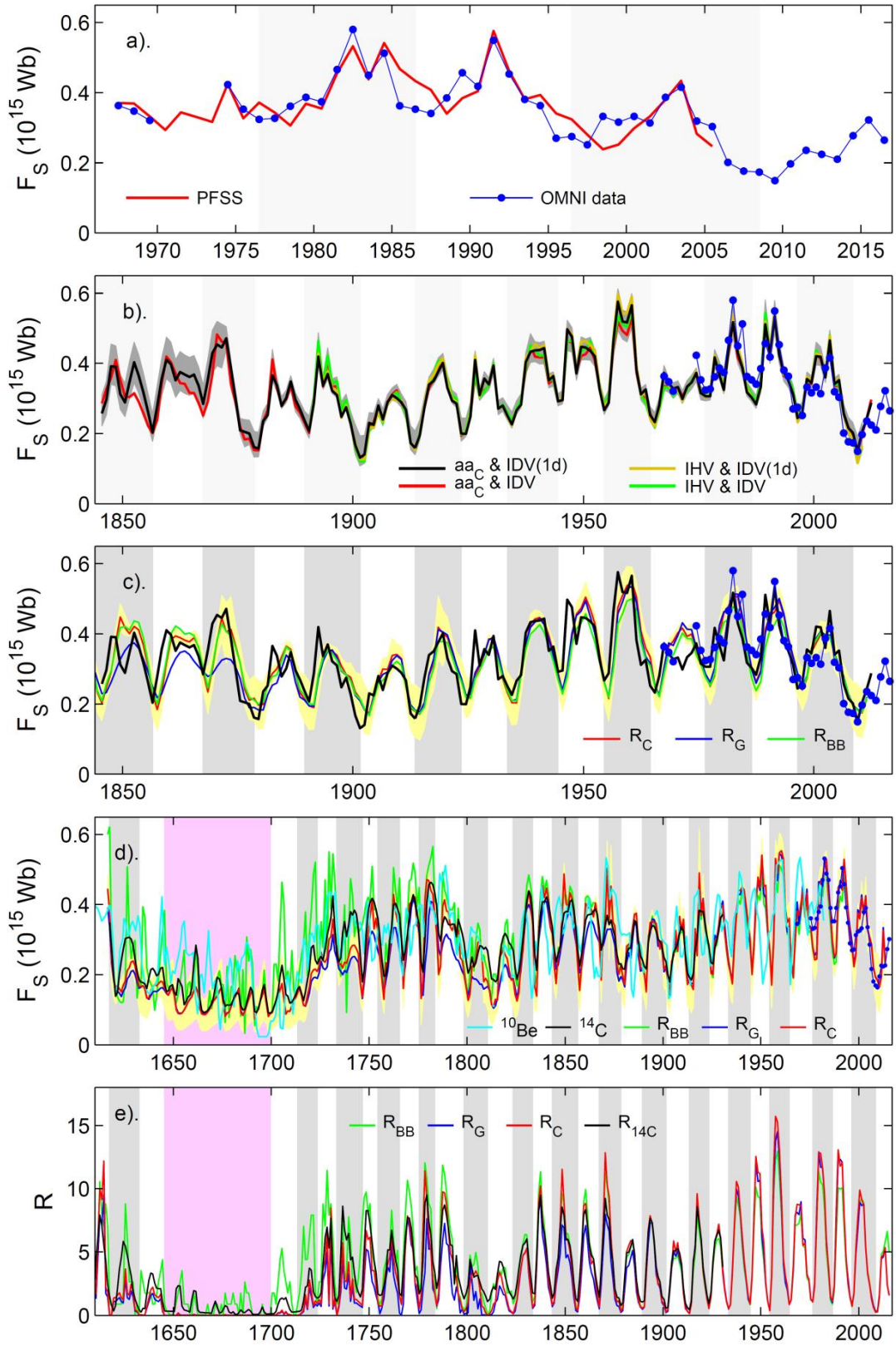

**Figure 1.** Variations of annual values of the signed open solar flux $F_S$, from various sources, showing how modern values (from magnetogram data via PFSS modelling and from the OMNI dataset of near-Earth interplanetary observations, panel a) are extrapolated back in time using geomagnetic observations (panel b), models based on sunspot numbers (panels c and d) and cosmogenic isotopes (panel d), Sunspot number sequences used as input to the models are shown in part e. Even-/odd-numbered solar cycles are shaded white and grey and the Maunder minimum is shaded pink. (See text for further details).

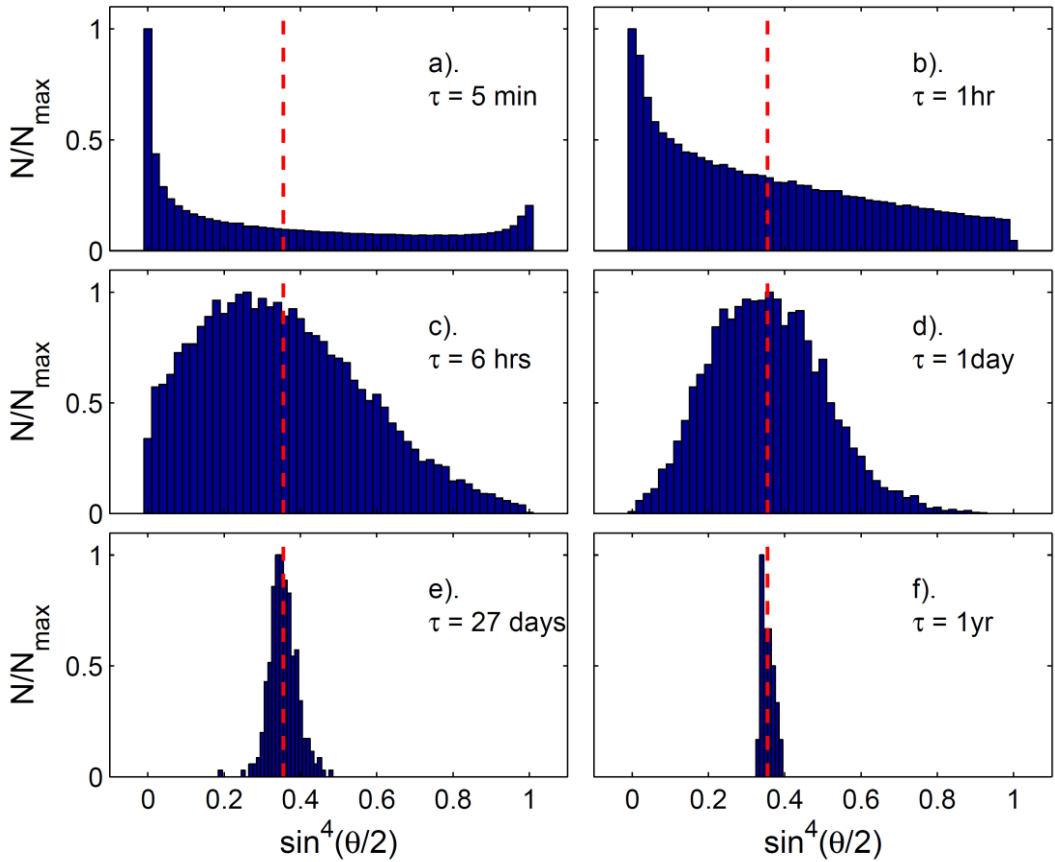

**Figure 2.** Distributions of the IMF orientation solar wind-magnetosphere coupling function factor, $\sin^4(\theta/2)$, for 1996-2016 (inclusive) for various averaging timescales, $\tau$, where $\theta = \tan^{-1}(|B_Y|/B_Z)$ where $B_Y$ and $B_Z$ are the Y- and Z-components of the IMF in the Geocentric Solar Magnetospheric (GSM) reference frame: (a). $\tau = 5$ min.; (b) $\tau = 1$hr.; (c) $\tau = 6$ hrs.; (d) $\tau = 1$ day; (e) $\tau = 27$ days; (f) $\tau = 1$ yr. $N$ is the number of samples in each $\sin^4(\theta/2)$ bin, the largest value of which for a given $\tau$ is $N_{max}$. In each panel the red dashed line is the overall mean for the whole dataset which, because the averaging intervals of length $\tau$ contain equal numbers of samples, equals the mean of each distribution (the "mean of means" or the "grand mean"). The raw $B_Y$ and $B_Z$ data used are of resolution 5 min. and have been rounded up/down to 2 decimal figures.

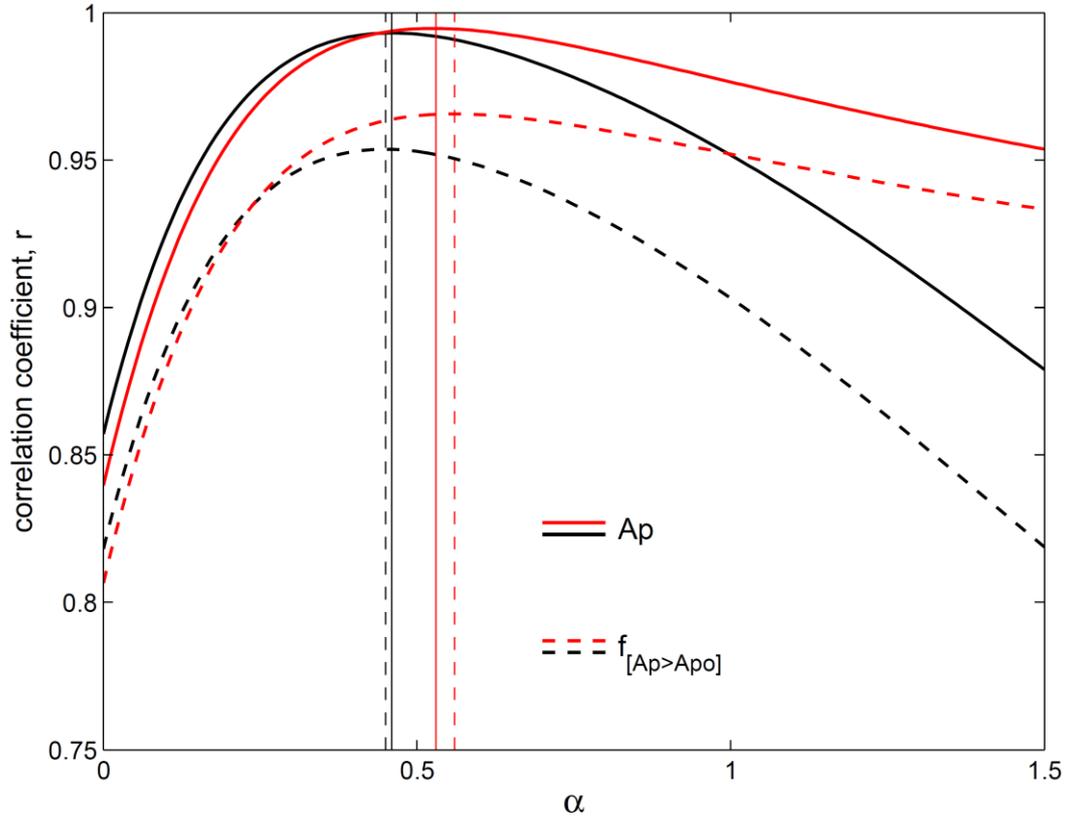

**Figure 3.** Correlation coefficients, *r*, as a function of α for the interval of near-continuous $P\alpha$ data (1996-2016). Solid lines are for correlations between annual $P\alpha$ and Ap and dashed lines are for correlations between $P\alpha$ and the fraction of days when Ap exceeds the overall 95 percentile level, $f_{[Ap>Apo]}$. The black lines use annual means ($\tau = 1yr$) of $P\alpha$ values computed from 5-minute means of interplanetary data, $<P\alpha>_{\tau=1yr}$ (i.e., the data are combined and then averaged) and the peak value (marked by the vertical solid black line) gives correlation A in Table 2. The red lines use $P\alpha_{ann}$ which is compiled using annual means of all available $N_{SW}$, $V_{SW}$, $B$ and $\sin^4(\theta/2)$ data (i.e., the data are averaged and then combined) and the peak values gives correlation B in Table 2. The 95-percentile of the overall distribution of all Ap values (observed since the data series began in 1932) is Apo = 36nT.

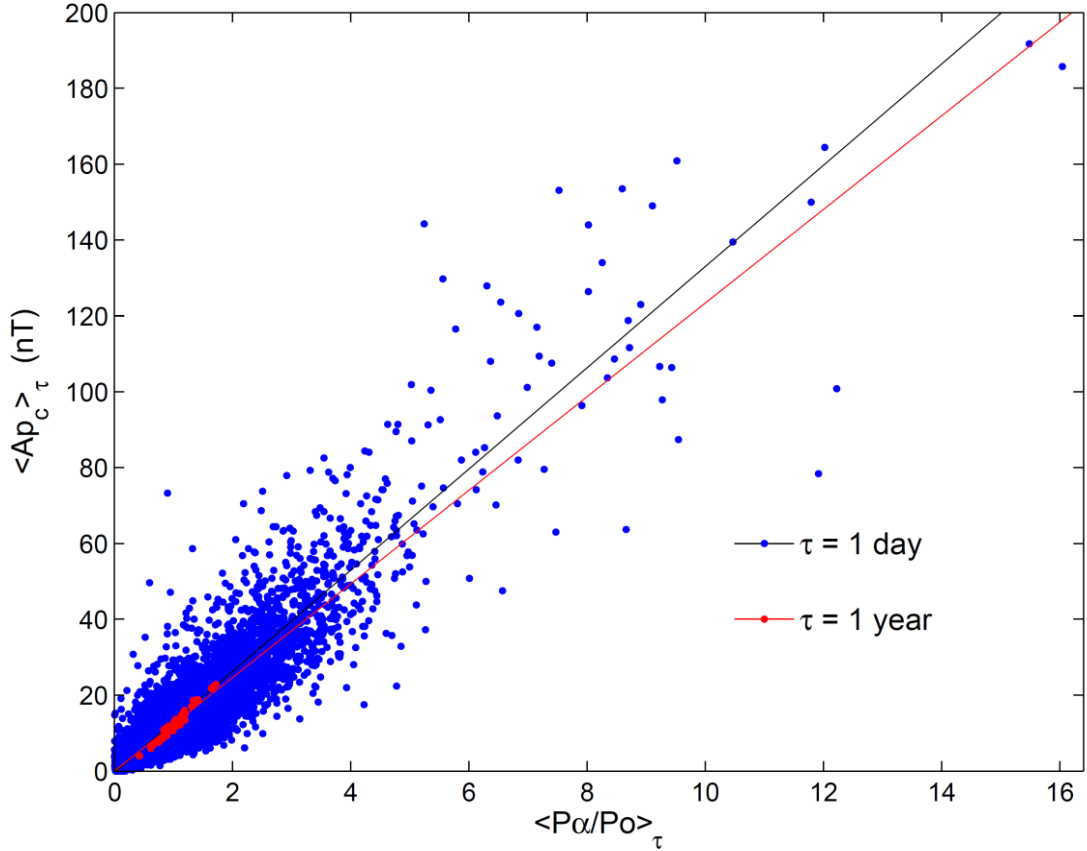

**Figure 4.** Scatter plots of $Ap_c$ as a function of normalised $P\alpha$ values (for the best fit $\alpha$ of 0.46) based on 1-hour resolution interplanetary data for 1964-2016. $Po$ is the average of $P\alpha$ for all the available data from 1964-2016, and $Ap_c$ are hourly Ap values (linearly interpolated from the observed 3-hourly Ap values) that are coincident with a valid hourly $P\alpha$ value. The blue dots are for daily means ($\tau = 1$ day) and the red dots for annual means ($\tau = 1$ year). The black line is the linear regression fit to the $\tau = 1$ day data and gives the correlation C in Table 2, the red line is the linear regression fit to the $\tau = 1$ year data and gives the correlation D in Table 2.

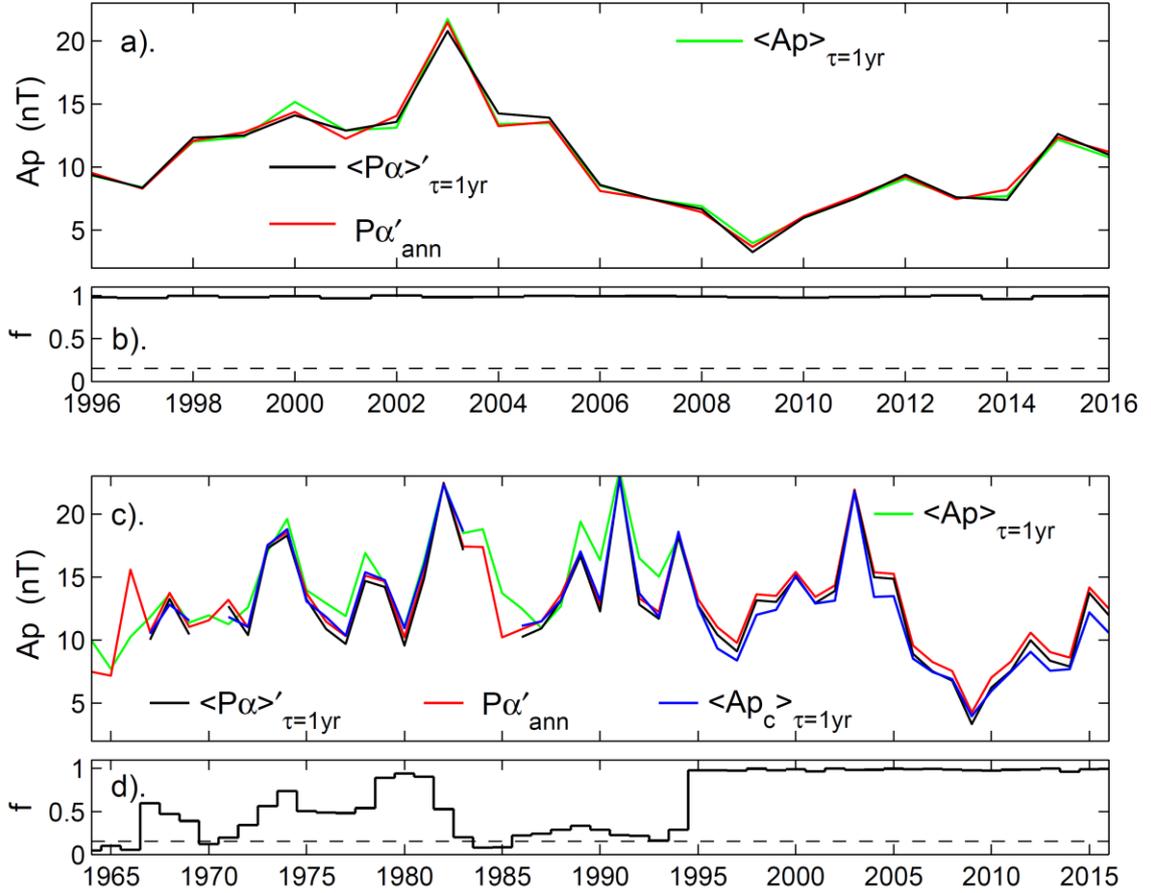

**Figure 5.** Time series of observed Ap and best-fits to Ap using $P\alpha$ estimates, both using annual averaging intervals. The fitted series use linear regressions and are denoted by a prime: tests for homoskedacity, normal distribution of residuals, linearity are all readily passed because correlation coefficients are so high. (a) and (b) are for 1996-2016, (c) and (d) for 1964-2016. Panels (b) and (d) show the data availability, giving the fraction $f$ of hours for which an hourly mean value of $P\alpha$ is available. In (a) and (c) the green line gives the means of all the Ap data, $<Ap>_{\tau=1yr}$ and the blue line gives the means of only the Ap data that are coincident with valid $P\alpha$ data $<Ap_c>_{\tau=1yr}$. (Note that in (a) the blue line is essentially identical to the green and so is completely overwritten by it). In (a) the black line uses annual means ($\tau = 1yr$) of $P\alpha$ values computed from 5-minute means of interplanetary data, $<P\alpha>'_{\tau=1yr}$ (i.e., the data are combined and then averaged) and is based on the correlation A in Table 2. The red line uses $P\alpha_{ann}$ which is compiled using annual means of all available $N_{SW}$, $V_{SW}$, $B$ and $\sin^4(\theta/2)$ data (i.e., the data are averaged and then combined) and is based on correlation B in Table 2. In panel (d) the black line uses $<P\alpha>_{\tau=1yr}$ and $<Ap_c>_{\tau=1yr}$ and is based on correlation E in Table 2 whereas the red line uses $P\alpha_{ann}$ and $<Ap_c>_{\tau=1yr}$ and is based on correlation D in Table 2.

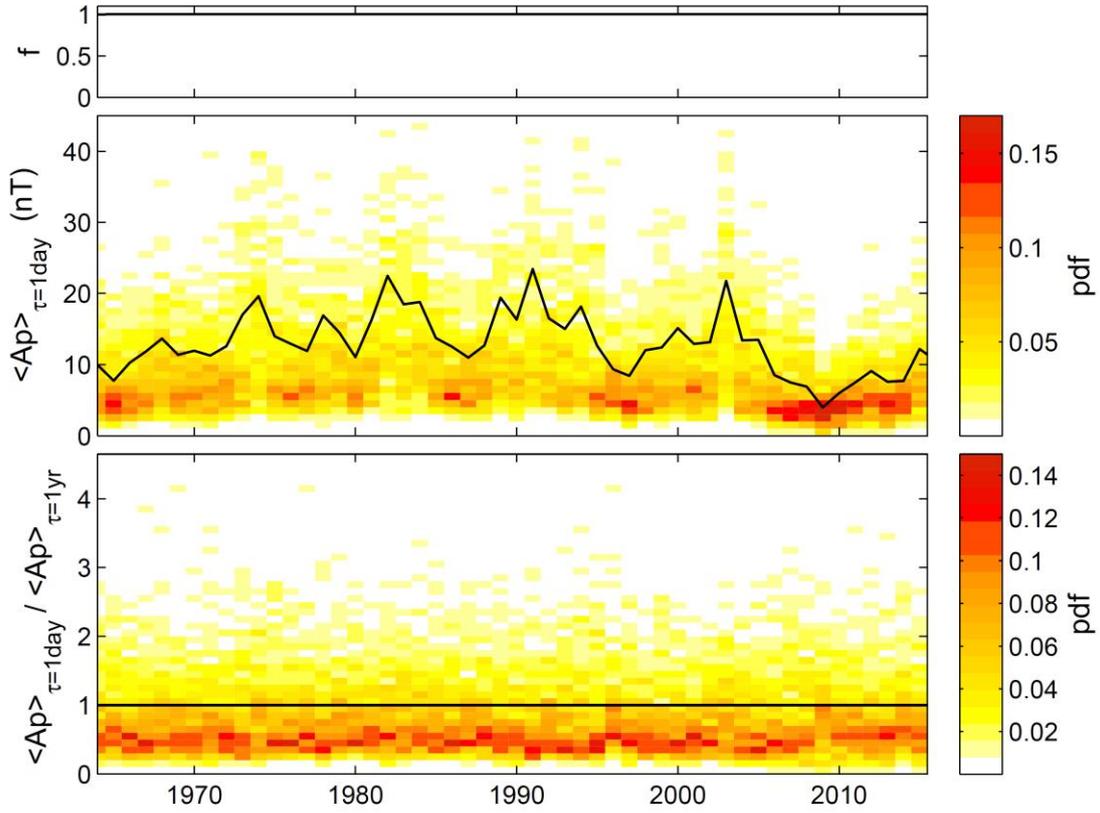

**Figure 6.** Analysis of Ap means distributions over 1964-2016. (Top) The fraction of time in each year, $f$, for which a valid Ap is available. (Middle) Annual probability distribution function (pdfs) of daily means Ap ($<Ap>_{\tau=1day}$) are colour coded and the black gives the annual means of Ap, $<Ap>_{\tau=1yr}$. Probabilities are evaluated in bins of Ap that are 1 nT wide. (Bottom) Annual probability distribution function (pdfs) of normalised daily Ap, $<Ap>_{\tau=1day}/<Ap>_{\tau=1yr}$, are colour coded and the black gives the annual means of this normalised Ap which is unity. Probabilities are evaluated in bins of normalised Ap that are 0.1 wide.

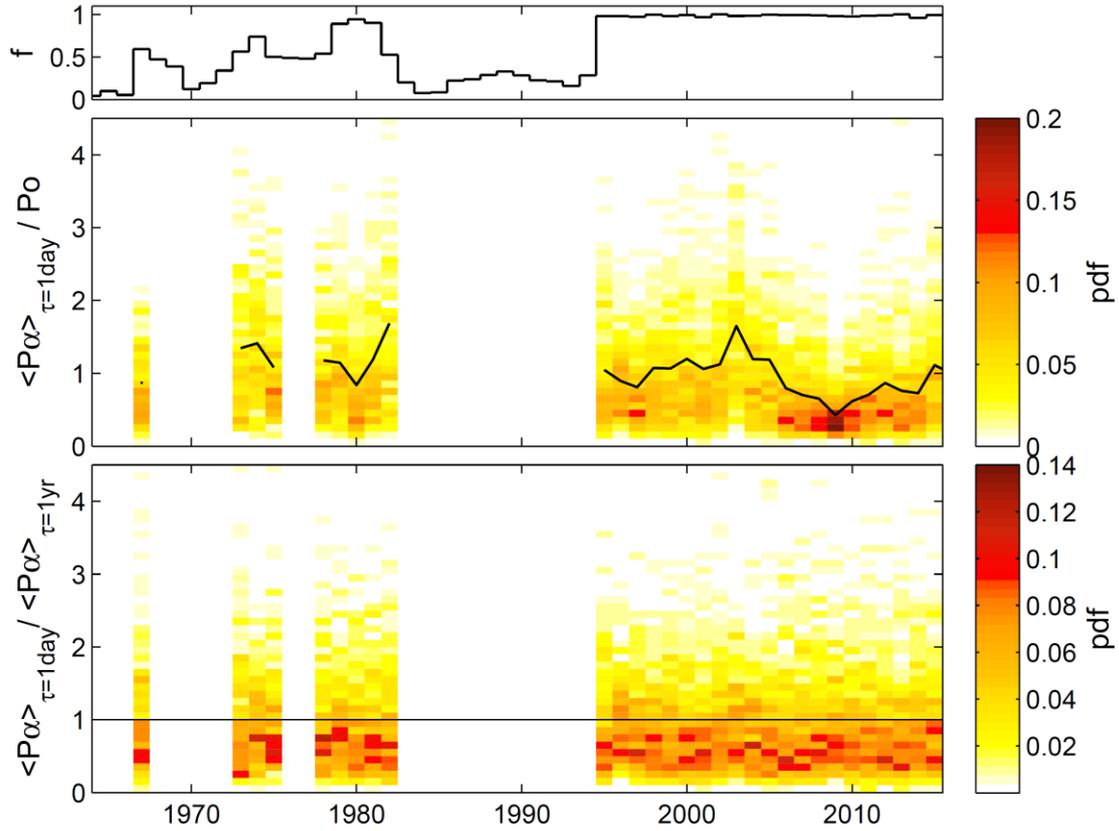

**Figure 7.** Same as figure 5 for the estimated power input into the magnetosphere, $P\alpha$. (Top) The fraction of time in each year, $f$, for which a valid $P\alpha$ value is available. (Middle) Annual probability distribution function (pdfs) of daily means of $P\alpha/Po$ ($<P\alpha>_{\tau=1day}/Po$) are colour coded and the black line gives the annual means of $<P\alpha>_{\tau=1yr}/Po$. Probabilities are evaluated in bins of $P\alpha/Po$ that are 0.1 wide and years for which $f < 0.5$ are omitted. (Bottom) Annual probability distribution function (pdfs) of normalised daily $P\alpha$, $<P\alpha>_{\tau=1day}/<P\alpha>_{\tau=1yr}$, are colour coded and the black gives the annual means of this normalised $P\alpha$ which is unity. Probabilities are evaluated in bins of normalised $P\alpha$ that are 0.1 wide.

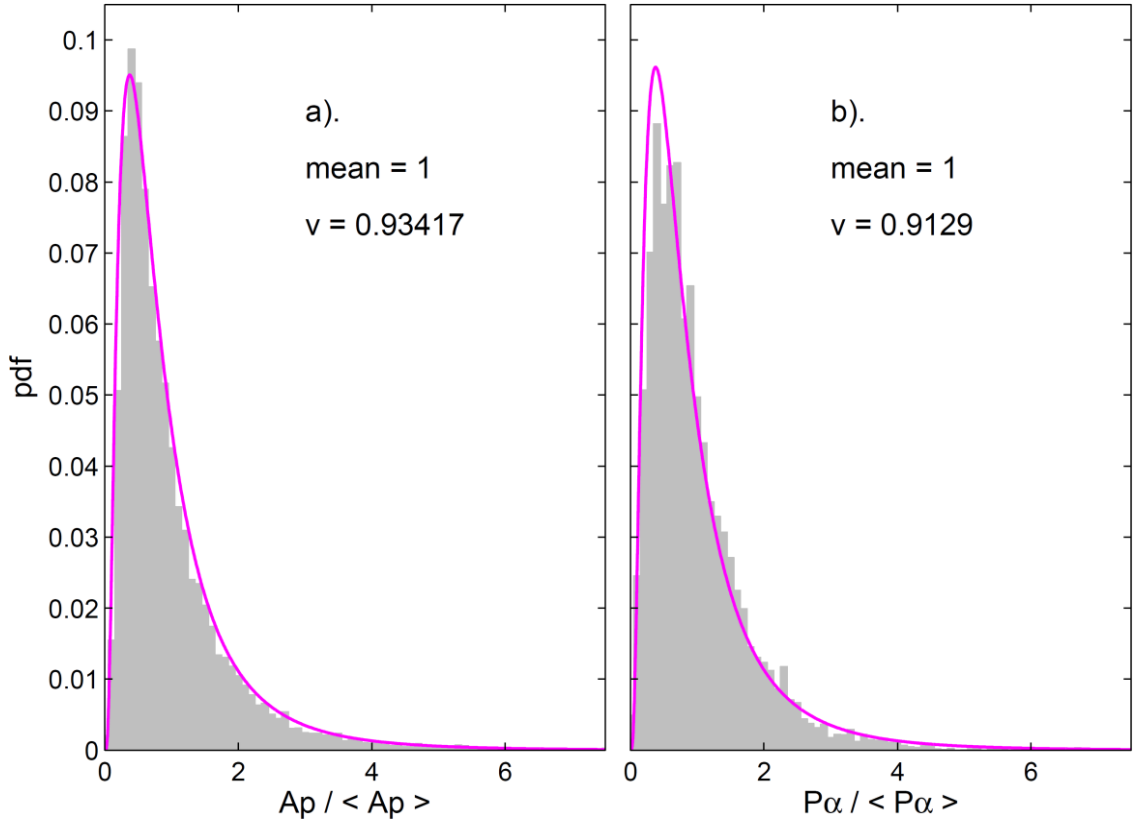

**Figure 8.** (Grey histograms) Average pdfs for all years of (a) $<Ap>_{\tau=1day}/<Ap>_{\tau=1yr}$ and (b) $<P\alpha>_{\tau=1day}/<P\alpha>_{\tau=1yr}$. In each case the best-fit lognormal distribution is shown, with mean of unity (by definition) and a variance $v = 0.934$ in (a) and $v = 0.913$ in (b). Distributions for individual years (not shown) are a bit noisier, particularly for $<P\alpha>_{\tau=1day}/<P\alpha>_{\tau=1yr}$ in years when data availability *f* is low, but give very similar best fit *v* values, as indicated by the horizontal uniformity of the bottom panels of figures (5) and (6): for the annual normalised Ap distributions, the mean ± 1 standard deviation of the fitted *v* values for the 53 years is 0.934±0.031; for the annual normalised $P\alpha$ distributions, *v* values for the 31 years with *f* > 0.5 is 0.915±0.025.

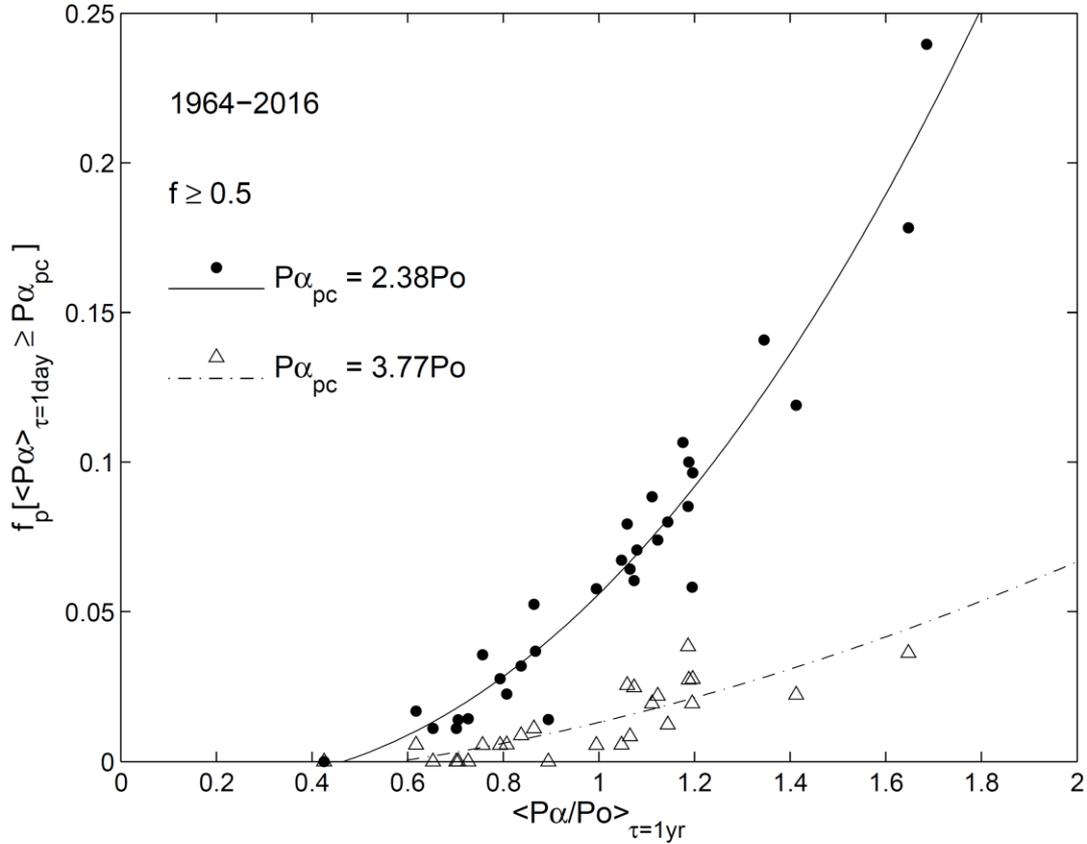

**Figure 9.** Variation of the fraction $f_p$ of daily means of $P\alpha$ in one year that are in the top 5% (solid points) and top 1% (open triangle) of all the 12198 available daily means of $P\alpha$ for the interval 1964-2016, shown as a function of the annual mean ($<P\alpha>_{\tau=1yr}/Po$). These data use the optimum $\alpha$ of 0.53. The annual fraction $f_p[<P\alpha>_{\tau=1day} \geq P\alpha_{pc}]$ is shown for two fixed thresholds: $P\alpha_{pc} = 2.38Po$ (which defines the 95$^{th}$ percentile of the cumulative probability distribution of all available $<P\alpha>_{\tau=1day}$ values from the interval 1964-2016 which number 12198); and $P\alpha_{pc} = 3.77Po$ (which defines the 99$^{th}$ percentile of the same cdf). Points for these two thresholds are shown by solid circles and open triangles, respectively. Only the 31 years in which data availability $f$ exceeds 50% are shown which account for 10186 samples out of the available total of 12198. (The other years show the same trend but the smaller number of samples increases the scatter). The solid line is a third order polynomial fit to the data points for $P\alpha_{pc} = 2.38Po$ and the dashed line is that for $P\alpha_{pc} = 3.77Po$. Both reveal that $fp$ shows an increasing trend with the annual mean value, but it is not, in general, linear. The smaller numbers associated with higher percentiles of the cdf mean that the scatter is increased.

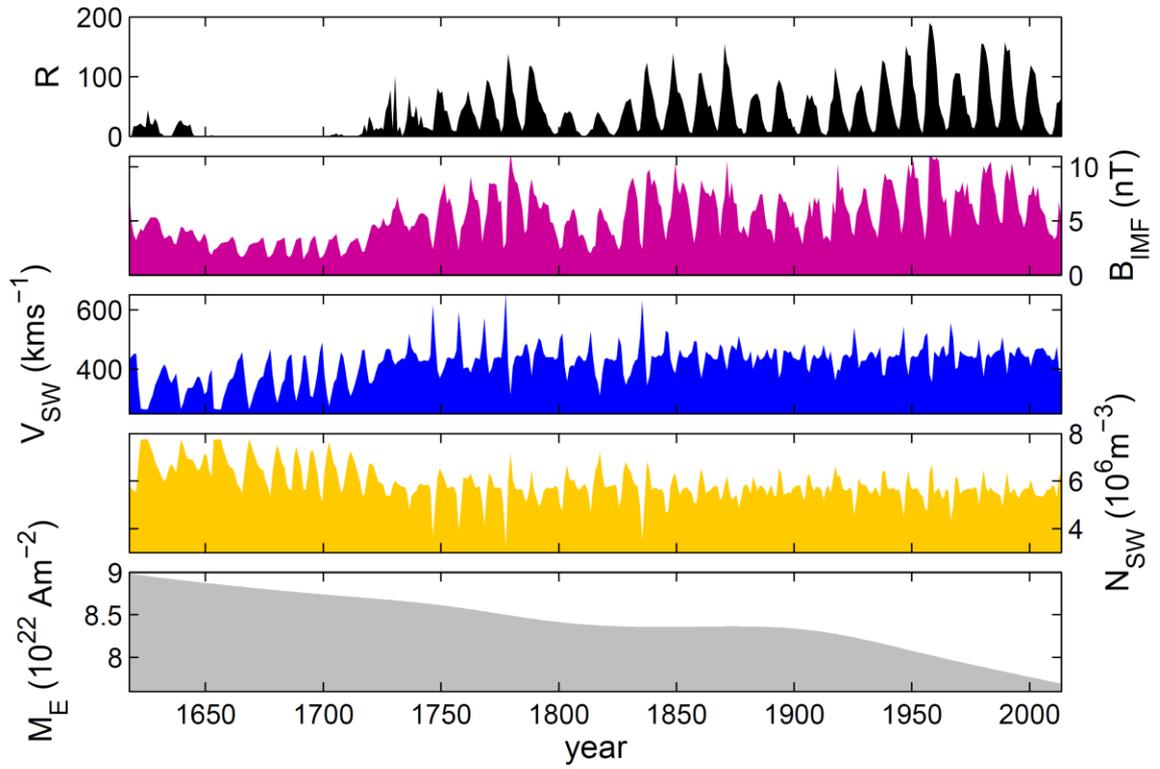

**Figure 10.** Time series of modelled annual values on near-Earth interplanetary parameters over 1612-2015 by Owens et al. (2017): (a) the input sunspot number data series, $R_C$; (b) the IMF field strength, $B$; (c) the solar wind speed, $V_{SW}$; (d) the solar wind number density, $N_{SW}$. Panel (e) shows the Earth's magnetic dipole moment, $M_E$, from a spline of the gufm1 (for before 1900) and IGRF (for after 1900) models.

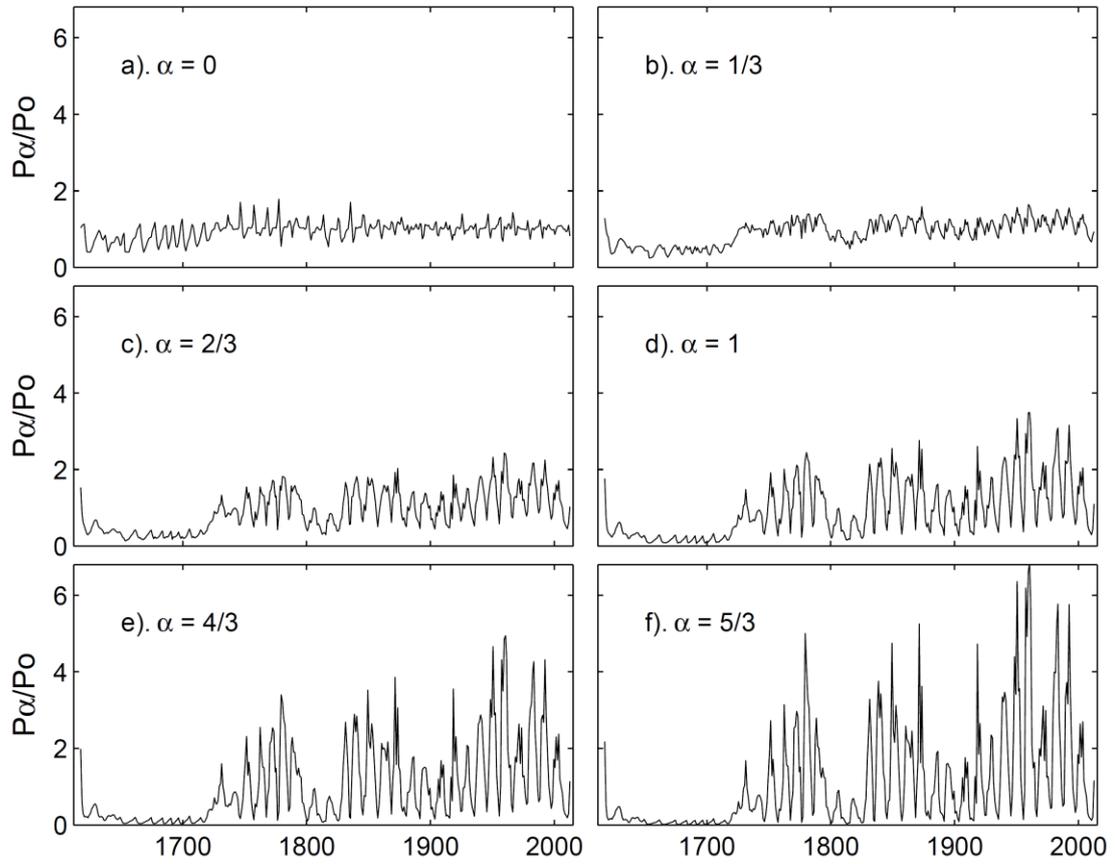

**Figure 11.** Predicted variations of annual $P\alpha/Po$ values for six values of the coupling exponent $\alpha$: (a) $\alpha = 0$; (b) $\alpha = 1/3$; (c) $\alpha = 2/3$; (d) $\alpha = 1$; (e) $\alpha = 4/3$; and (f) $\alpha = 5/3$.

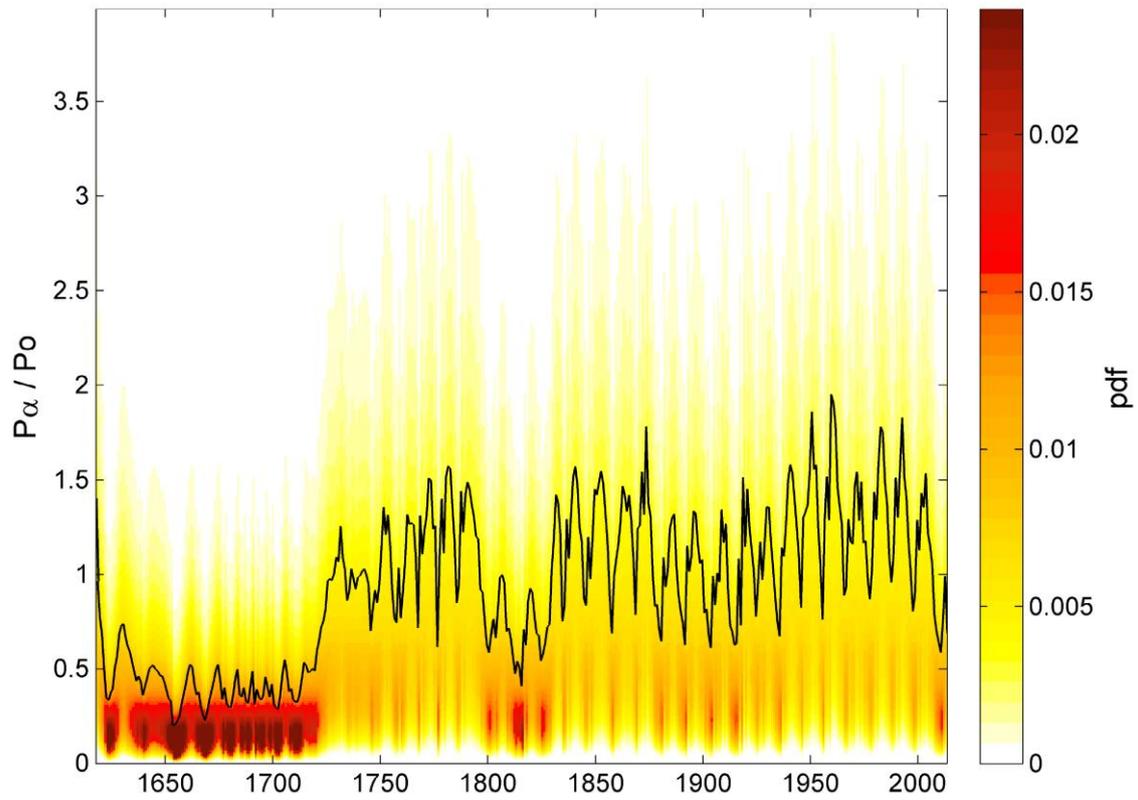

**Figure 12.** Annual pdfs and mean values of the normalised power input to the magnetosphere, $P\alpha/Po$ for 1612-2015 and for the best-fit coupling exponent for the Ap index of 0.53. The colour contours give the pdf, evaluated for bins in $P\alpha/Po$ that are of width 0.01. The $Po$ used here is the average value of $P\alpha$ from spacecraft measurements over 1964-2016 (inclusive).